\documentclass[letterpaper]{JHEP3}

\usepackage{subfigure}
\usepackage{graphicx}
\usepackage{amsmath, amssymb}

\newcommand{\eq}{\begin{equation}}
\newcommand{\eqe}{\end{equation}}

\newcommand{\om}{\omega}
\newcommand{\si}{\sigma}
\newcommand{\la}{\lambda}
\newcommand{\de}{\delta}
\newcommand{\eqa}{\begin{eqnarray}}
\newcommand{\eqae}{\end{eqnarray}}

\newcommand{\e}{\epsilon}
\newcommand{\braket}[1]{\langle #1 \rangle}
\newcommand{\beqs}{\begin{equation*}}
\def\beq{\begin{equation}}
\newcommand{\eeqs}{\end{equation*}}
\def\eeq{\end{equation}}
\newcommand{\beqas}{\begin{eqnarray*}}
\newcommand{\beqa}{\begin{eqnarray}}
\newcommand{\eeqas}{\end{eqnarray*}}
\newcommand{\eeqa}{\end{eqnarray}}
\newcommand{\al}{\alpha}

\newcommand{\Ga}{\Gamma}

\title{Dualities for Loop Amplitudes of $\mathcal{N}=6$ Chern-Simons Matter Theory.\;}
\author{Wei-Ming Chen\footnote{Email: tainist@gmail.com}$^{~1,2}$, and Yu-tin Huang\footnote{Email: yhuang@physics.ucla.edu}$^{~3,4}$
\\ \\ \\
\it $^1$ Department of Physics and Center for Theoretical Sciences,\\ National Taiwan University,\\
Taipei 10617, Taiwan, R.O.C\\
\\
\it $^2$ California Institute of Technology,\\
Pasadena, CA 91125, USA
\\
\\
\it $^3$ Department of Physics and Astronomy,\\ UCLA,\\
Los Angeles, CA 90095-1547, USA\\
\\
\it $^4$ Kavli Institute for Theoretical Physics,\\  University of California at Santa Barbara,\\
CA 93106-4030, USA}
\abstract{ In this paper we study the one- and two-loop corrections to the four-point amplitude of $\mathcal{N}=6$ Chern-Simons matter theory. Using generalized unitarity methods we express the one- and two-loop amplitudes in terms of dual-conformal integrals. Explicit integration by using dimensional reduction gives vanishing one-loop result as expected, while the two-loop result is non-vanishing and matches with the Wilson loop computation. Furthermore, the two-loop correction takes the same form as the one-loop correction to the four-point amplitude of $\mathcal{N}=4$ super Yang-Mills. We discuss possible higher loop extensions of this correspondence between the two theories. As a side result, we extend the method of dimensional reduction for three dimensions to five dimensions where dual conformal symmetry is most manifest, demonstrating significant simplification to the computation of integrals.  \;   
}
\preprint{ UCLA-TEP-11-109, NSF-KITP-11-129}
\begin{document}

\section{Introduction}

In the past year there has been much advancement in the study of perturbative S-matrix of $\mathcal{N}=6$ Chern-Simons matter theory  constructed by Aharony, Bergman, Jafferis and Maldacena (ABJM)~\cite{Aharony:2008ug}. It first was noticed in~\cite{Bargheer:2010hn} that the four- and six-point tree amplitude of this theory exhibit an $OSp(6|4)$ Yangian symmetry. The generators of this Yangian were later reformulated as superconformal generators in a dual space, which consists of three space-time, three R-symmetry space and six fermionic coordinates~\cite{Huang:2010qy}. Using a newly constructed recursion relation for tree amplitudes, it was shown that the symmetry of the four- and six-point amplitude is extended to arbitrary point, and to the cut constructible part of the loop integrand~\cite{Gang:2010gy}. In parallel development, it was found that the tree-level amplitudes can be reproduced by an orthogonal Grassmannian integral, which can be shown to exhibit the same Yangian invariance~\cite{Lee:2010du}.

All of the novel features for ABJM amplitudes discussed above are shared with $\mathcal{N}=4$ super Yang-Mills~\cite{ DualConformal0,DualConformal, Grassmanian} and it is natural to argue that the origin of these common features, in particular the presence of dual conformal symmetry, lies in the fact that both theories are dual, via $AdS$/CFT, to a string theory where these symmetries are just the isometries of the dual background. For $\mathcal{N}=4$ super Yang-Mills, dual superconformal, and hence the $SU(2,2|4)$ Yangian symmetry, indeed can be understood as the self-duality of the $AdS_{5}\times S^5$ background under fermionic T-duality~\cite{FermiT}. However, there are reasons to suspect that dual conformal symmetry may not be entirely tied to $AdS$/CFT. In particular, dual conformal symmetry can be extended to tree-level amplitudes and loop integrands of maximally supersymmetric Yang-Mills in higher dimensions, where the gauge theory is not classically conformal~\cite{Dennen:2010dh,CaronHuot:2010rj}.  Indeed while ABJM is dual to type IIA string theory on $AdS_{4}\times \mathbf{CP}^3$,  attempts in demonstrating the self-duality for the corresponding sigma model has failed so far~\cite{FermiT2}. 

A related issue is that the existence of both the original and dual superconformal symmetry for the amplitude, implies that there exists a dual object for which the dual symmetries are local. For $\mathcal{N}=4$ super Yang-Mills, the dual object was proposed to be a (super-)Wilson loop~\cite{Alday:2007hr,WilsonLoop1,Brandhuber:2007yx, WilsonLoop2, WilsonLoop3} and later promoted to correlation functions with light-like separations~\cite{Operator1, Operator2}. The situation for ABJM is again unclear due to the fact that the dual objects proposed for $\mathcal{N}=4$ super Yang-Mills are defined chirally. More precisely, the $n$-cusps super-Wilson loop has homogeneous Grassmann degree for all $n$, and the degree depends on the helicity structure of the amplitude which the particular Wilson loop is dual to. Since the $n$-point amplitude for ABJM has Grassmann degree $n3/2$, the lack of homogeneous Grassmann degree implies that the (super-)Wilson loop/amplitude duality, if it exists, should be structurally different than the duality for $\mathcal{N}=4$ super Yang-Mills.   

The purpose of the paper is to take initial steps in studying the possibility of Wilson loop/amplitude duality for ABJM theory. We will focus on the four-point amplitude where the lack of chirality is not an issue, since the amplitude depends on the Grassmann variables only through a trivial supermomenta delta function, which is not expected to be captured by the Wilson loop. The one-loop four-point amplitude has been shown to vanish in~\cite{Agarwal:2008pu}, and this is in agreement with the one-loop four-cusp Wilson loop computation~\cite{NewWilson,Henn:2010ps}. However, in~\cite{Henn:2010ps} it was shown that the Wilson loop at one-loop vanishes for both Chern-Simons and ABJM theory, while at two-loop for both does not vanish and gives different results. Since Chern-Simons theory has trivial S-matrix, the non-vanishing two-loop result rules out Wilson loop/amplitude duality for this theory. Thus this indicates the matching of one-loop result is insufficient to establish Wilson loop/amplitude duality for Chern-Simons-like theory, and a two-loop computation would be necessary for any non-trivial statements.\footnote{On the other hand, recently it was also shown~\cite{Bianchi:2011rn} that for $\mathcal{N}=2,3,6,8$ Chern-Simons matter theory, the one-loop $n$-point correlator divided by its tree-level expression coincides with a light-like $n$-polygon Wilson loop at the integrand level. Hence correlator-Wilson loop duality has non-trivial evidence at one-loop level. }

We proceed by utilizing the fact that the integrand, to all loop orders, is dual conformal covariant when the integrand is defined in three-dimensions. This allows us to write down a basis for the integrand which consists of dual conformal integrals. The relevant coefficients of these integrals are then fixed via  generalized unitarity method~\cite{UnitarityMethod}(for a recent review see \cite{Bern:2011qt}). At one-loop, the only four-point dual conformal integral is a tensor integral. Using the $s$ and $t$-channel cuts we show that this integral is indeed the correct integrand, which vanishes up to order $\mathcal{O}(\epsilon)$ upon integration. This resolves the paradox that while the one-loop amplitude is expected to vanish, its unitarity cut does not vanish manifestly.\footnote{In fact it can be shown that the the two-particle cut vanishes upon phase space integration~\cite{Agarwal:2008pu}.}

\begin{figure}[h]\centering
\hspace{-4mm}\includegraphics[scale=0.76]{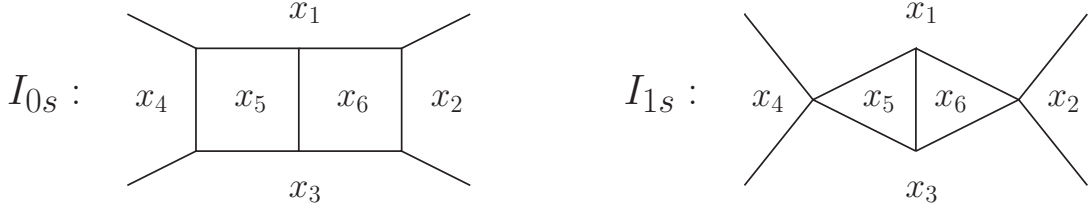}
\caption{Integrals that contribute to the two-loop amplitude. Here only the propagators and the positions of the dual space coordinates are shown. The numerator for each integral is chosen such that the it is invariant under conformal symmetry in dual space.}
\label{sum}
\end{figure}

At two-loop, the integrand is given by two dual conformal integrals:
\eq
\mathcal{A}_4^{2-loop}=\left(\frac{N}{K}\right)^2\mathcal{A}_4^{tree}\left[-I_{0s}+I_{1s}+\left(s\rightarrow t\right)\right],
\eqe
where $I_{0s}$ is a double box integral with tensor numerators and $I_{1s}$ is a double triangle ``kite" integral. The propagator structures are given in fig.\ref{sum}. The notation $s\rightarrow t$ corresponds to the inclusion of integrals that are cyclic rotations ($1\rightarrow4$, $2\rightarrow1$, e.t.c) of $I_{0s}$, $I_{1s}$, and the integrals written in five-dimensional notation, are given as 
\eqa
 I_{0s}&\equiv& \int \mathcal{D}^3 X_5\mathcal{D}^3X_6 \frac{16\epsilon(5,1,2,3,4)\epsilon(6,1,2,3,4)}{X_{51}^2X_{53}^2X_{54}^2X_{56}^2X_{61}^2X_{63}^2X_{62}^2X_{42}^2},\\
 I_{1s}&\equiv& \int \mathcal{D}^3X_5\mathcal{D}^3X_6 \frac{X_{31}^2}{X_{51}^2X_{53}^2X_{56}^2X_{61}^2X_{63}^2},
\eqae
where the $X_{ij}$s are the difference of three-dimensional dual space coordinates conformally embedded in five dimensions, $\epsilon(i,j,k,l,m)\equiv \epsilon_{\mu\nu\rho\sigma\tau}X_i^\mu X_j^\nu X_k^\rho X_l^\sigma X_m^\tau$, and $\mathcal{D}^3X_5\mathcal{D}^3X_6$ are the projective integration measures. The precise embedding of the three-dimensional kinematics in five-dimensional represented will be discussed in the paper, one only needs to note that the relation to the three-dimensional momenta are given by $-2X_i\cdot X_{i+1}=p^2_i$. The two integrals are manifestly infrared power safe, which can be seen from the fact that while massless corners appear in $I_{0s}$, the numerator of $I_{0s}$ vanishes in the collinear region of the corners.\footnote{We thank Simon Caron-Hout for pointing this out.} 

Explicit integration has been carried out using dimensional reduction, where the tensor algebra are done in strictly three dimensions to obtain scalar integrands, and then one analytically continues to $D=3-2\epsilon$. This regularization scheme has been successfully applied to Chern-Simons like theories~\cite{Chen:1992ee} to three loop order, and was employed for the Wilson-loop computation as well~\cite{Henn:2010ps}. The result is given by 
\eq
\framebox[15cm][c]{$\displaystyle\mathcal{A}_4^{2-loop}=-\frac{1}{16\pi^2}\left(\frac{N}{K}\right)^2\mathcal{A}_4^{tree}\left[\frac{(-s/\tilde \mu^2 )^{-2\epsilon}}{(2\epsilon)^2}+\frac{(-t/\tilde \mu^2 )^{-2\epsilon}}{(2\epsilon)^2}-\frac{1}{2}\log^2\left(\frac{-s}{-t}\right)+a+\mathcal{O}(\epsilon)\right]$}~,
\eqe
where $a=-8.02109\pm 6.57\times 10^{-5}$ and $\tilde\mu^2\equiv \left(e^{-\gamma_E}8\pi\right)\mu^2$. This is indeed the same form given by the two-loop Wilson loop computation in~\cite{Henn:2010ps}, 
\eq
\langle W_4\rangle^{ABJM}=1-\left(\frac{N}{K}\right)^2\left[\frac{(-\mu'^2 x_{13}^2)^{-2\epsilon}}{(2\epsilon)^2}+\frac{(-\mu'^2 x^2_{24})^{-2\epsilon}}{(2\epsilon)^2}-\frac{1}{2}\log^2\left(\frac{-x^2_{13}}{-x^2_{24}}\right)+const+\mathcal{O}(\epsilon)\right]\,.
\eqe
Here, $N$ is the rank of the gauge group $U(N)\times U(N)$ and $K$ is the Chern-Simons level. Thus not only those the two match in degree of IR-divergence, the finite term matches up to a constant as well. This provides the first non-trivial evidence of Wilson loop/amplitude duality for ABJM theory.  Interestingly that while both the $t$- and $s$-channel cuts vanish, since one factors into two odd-point tree amplitudes while the other factors into a one-loop and a tree, the requirement of consistency in both cuts gives a non-vanishing two-loop amplitude.

An interesting feature of this result, as first noted in~\cite{Henn:2010ps}, is that the two-loop ABJM amplitude takes on striking similarity with the one-loop four-point amplitude of $\mathcal{N}=4$ super Yang-Mills\cite{Green:1982sw}:
\eq
\mathcal{A}_4^{\mathcal{N}=4}=\frac{ig^2N}{8\pi^2}\mathcal{A}_4^{tree}\left[-\frac{(-s/\mu^2)^{-\epsilon}}{\epsilon^2}-\frac{(-t/\mu^2 )^{-\epsilon}}{\epsilon^2}+\frac{1}{2}\log^2\left(\frac{-s}{-t}\right)+4\zeta_2+\mathcal{O}(\epsilon)\right]\,.
\eqe
One sees that again up to a constant piece, the two differ by changing $\epsilon\rightarrow 2\epsilon$ which reflects the difference between one- and two-loop. In fact, the similarity between two-loop ABJM and one-loop $\mathcal{N}=4$ super Yang-Mills four-point amplitudes can be argued on the grounds of dual conformal symmetry. Due to infrared singularities, the dual conformal symmetry of the planar integrand becomes anomalous and one can derive a corresponding anomalous Ward identity~\cite{Drummond:2007cf,Brandhuber:2009kh}. At four-point, this Ward identity is sufficient to fix the amplitude up to a constant due to the lack of dual conformal invariant cross-ratios. Since the infrared singularities of the two amplitudes match, this leads to matching  anomalous Ward identities, and thus forces the finite part of the four-point amplitude to match as well. The same result was observed on the Wilson loop side, where explicit analysis of the Wilson loop conformal Ward identity at two-loop for ABJM~\cite{Henn:2010ps} takes the same form as that of the one-loop result for $\mathcal{N}=4$ super Yang-Mills~\cite{Drummond:2007cf}. Since the anomalous Ward identity is controlled by the infrared singularity, the four-point amplitudes of the two theories will continue to match to higher loops provided that the infrared singularities match. This will have an immediate consequence in that the BDS ansatz~\cite{Bern:2005iz}, under minor modification, gives the all loop four-point amplitude of ABJM. We discuss some possibilities in sec.\ref{guess}.

We will also evaluate the integrals in an alternative way, where the tensor numerators are kept in their five-dimensional form and one continues to $D=3-2\epsilon$ only at the very end of the computation. This approach reduces the complexity of the evaluation significantly due to the fact that dual conformal invariance is manifest during the intermediate steps. In spirit, this is an extension of dimensional reduction since the tensors and their manipulation are kept in five dimensions. We will show that both at one- and two-loop this approach reproduces the conventional dimensional reduction result. 

After the publication of this paper, we are informed of another upcoming paper~\cite{Bianchi:2011dg} which computes the same two-loop four-point amplitude by using Feynman diagrams. The two results agree with each other, and \cite{Bianchi:2011dg} gives an analytic form of the constant $a=-4\zeta_2-3\ln^2 2$.

This paper is organized as follows:  In the next section we give a brief review of scattering amplitudes for ABJM theory. In particular we will discuss three-dimensional spinor-helicity formalism and dual superconformal symmetry of the tree-level amplitudes and integrand. In section~\ref{OneLoop}, we begin with the one-loop exercise. We will show that the unique dual conformal integral at four-point is the correct integrand by demonstrating that it satisfies the unitarity cuts. Explicit integration will be carried out using five-dimensional formulation which trivially vanishes. In section~\ref{TwoLoop}, we will construct a basis of dual conformal integrals, and we will fix the relative coefficients using unitarity cuts. In section~\ref{TL}, explicit computation is done for the ``kite" integral using integration by part techniques~\cite{Chetyrkin:1981qh}, while the tensor integrals are computed using Mellin-Barnes representation techniques. An introduction to Mellin-Barnes representation technique can be found in~\cite{Smirnov:2006ry}. We give the explicit details of the computation in the appendix. We will discuss possible relationship between ABJM and $\mathcal{N}=4$ super Yang-Mills amplitudes and end with the conclusion in section \ref{guess}. 
\section{ABJM Tree Amplitudes and Dual Superconformal Symmetry}
In this section we give a brief introduction to tree-level amplitudes of ABJM and three-dimensional spinor-helicity formalism~\cite{Bargheer:2010hn,Agarwal:2008pu,hl1,Agarwal:2011tz}. Spinor-helicity formalism allows one to express the amplitude in terms of independent on-shell variables. This becomes especially useful when one performs state sums for recursion relations and unitarity cut. Later we will discuss the hidden dual superconformal symmetry of the amplitude and its implication on loop integrand.  

In $SL(2,R)$ notation, a vector in three dimensions is given by a symmetric $2\times2$ matrix. The null condition then translate into the property that the matrix is rank one: 
\eq
p^{\alpha\beta} = p_{\mu} (\sigma^\mu)^{\alpha\beta} = \lambda^\alpha \lambda^\beta \,.
\eqe
where the indices $(\alpha,\beta)$ transform under $SL(2,R)$, and the $\lambda$s are defined up to a sign. The sign ambiguity corresponds to the invariance of a null vector under the three-dimensional little group, $Z_2$. Our conventions for spinors and gamma matrices are
summarized in appendix \ref{AppA1}. Lorentz invariants can be constructed by contracting the spinors using the $SL(2,R)$ metric
\eq
\langle ij\rangle \equiv \lambda_i^\alpha\lambda_{j\alpha}=\lambda_i^\alpha\epsilon_{\alpha\beta}\lambda_{j}^{\beta} \,,
\eqe
where $\epsilon_{12}=-\epsilon^{12}=1$. The vector and spinor Lorentz invariants are related by
\eq
(p_i+p_j)^2=2p_i\cdot p_j=-\langle ij\rangle^2\,.
\eqe

\subsection{ABJM Tree Amplitudes}
The S-matrix of Chern-Simons matter theory vanishes when one of the external lines is a gauge field, since the Chern-Simons gauge field does not carry any physical degrees of freedom. Thus the $n$-point amplitude is non-vanishing only if $n=2k$, $k\in N$. The non-vanishing amplitudes then consists of purely matter states which forms a vector representation under the R-symmetry. 

The ABJM theory~\cite{Aharony:2008ug} is a three-dimensional twisted Chern-Simons theory with bi-fundamental matter. The field content consists of four complex scalars, $(\phi^{A},\bar{\phi}_A)$, and four fermions, $(\psi_A,\bar{\psi}^A)$, where $A$ is the $SU(4)$ R-symmetry index. The matter fields transform in the bi-fundamental representation of $U(N)\times U(N)$ gauge symmetry carried by two $U(N)$ Chern-Simons gauge fields $A_{\mu}$ and $\hat{A}_{\mu}$. The explicit form of the action can be found in~\cite{action}. Since $\mathcal{N}=6$ is not maximal, the on-shell multiplet is contained in two superfields
\eqa
\nonumber\Phi(\eta)&=&\phi^4+\eta^I\psi_I+\frac{1}{2}\epsilon_{IJK}\eta^I\eta^J\phi^K+\frac{1}{3!}\epsilon_{IJK}\eta^I\eta^J\eta^K\psi_4,\\
\Psi(\eta)&=&\bar{\psi}^4+\eta^I\bar{\phi}_I+\frac{1}{2}\epsilon_{IJK}\eta^I\eta^J\bar{\psi}^K+\frac{1}{3!}\epsilon_{IJK}\eta^I\eta^J\eta^K\bar{\phi}_4,
\eqae
where the Grassmann odd variables $\eta^I$ carries $U(3)$ indices, i.e. $I=1,2,3$. Thus we see that only the $U(3)$ subgroup of the $SU(4)$ R-symmetry is manifest. Thus the bosonic variables $\lambda$ carries the kinematic information while the fermionic $\eta$s carries the information of the particle species of the external states. 

We will be interested in the color ordered amplitude. Color ordering can be straightforwardly defined for bi-fundamental theories and results in a color ordered amplitude $\mathcal{A}_n$ that is invariant under cyclic permutation of two sites~\cite{Bargheer:2010hn}:
\eq
\mathcal{A}_n(1,2,\cdot\cdot\cdot, n)=(-)^{\frac{(n-2)}{2}}\mathcal{A}_n(3,\cdot\cdot\cdot,n,1,2).
\eqe
Invariance under the $U(1)$ of $U(3)$ restricts an $n$-point ABJM superamplitude to be of Grassmann degree $3n/2$. In particular, the four- and six-point superamplitude is given by~\cite{Bargheer:2010hn,Gang:2010gy}:
 \eqa
 \nonumber \mathcal{A}^{tree}_4(1,2,3,4)&=&i\frac{\delta^3(P)\delta^6(Q)}{\langle21\rangle\langle14\rangle}=-i\frac{\delta^3(P)\delta^6(Q)}{\langle23\rangle\langle34\rangle},\\
 \mathcal{A}^{tree}_6(1,2,3,4,5,6)&=& \frac{i\delta^3(P)\delta^6(Q)}{(p_{123})^2} \left[\frac{ \left( \epsilon_{\bar{i}\bar{j}\bar{k}} \langle \bar{i} \bar{j} \rangle \eta_{\bar{k}}^I+ i \,\epsilon_{ijk} \langle ij \rangle \eta_k^I  \right)^3}
{(\langle 1 | p_{123}| 4 \rangle - i \langle 23 \rangle \langle 56 \rangle )( \langle 3 | p_{123} | 6 \rangle - i \langle 12 \rangle \langle 45 \rangle )} \right. \nonumber
\\
&&\hspace{1cm}
\left.-\frac{ \left( \epsilon_{\bar{i}\bar{j}\bar{k}} \langle \bar{i} \bar{j} \rangle \eta_{\bar{k}}^I - i \,\epsilon_{ijk} \langle ij \rangle \eta_k^I  \right)^3}
{(\langle 1| p_{123} |4 \rangle +i \langle 23 \rangle \langle 56 \rangle )(  \langle 3|p_{123} |6 \rangle  + i \langle 12 \rangle \langle 45 \rangle )}\right],
\label{ABJMTree}
 \eqae
 where ($i,j,k=1,2,3$) and ($\bar{i},\bar{j},\bar{k}=4,5,6$), and 
 \eq
 \delta^3(P)=\delta^3\left(\sum^n_ip_i\right),\;~\delta^6(Q)=\prod^3_{I=1}\delta\left(\sum^n_i \lambda^\alpha_i\eta^I_i\right)\delta\left(\sum^n_i \lambda_{i\alpha}\eta^I_i\right).
 \eqe

\subsection{Dual Superconformal Symmetry }
The dual symmetry is an $OSp(6|4)$ superconformal symmetry acting on a ``dual space" parameterized by the coordinates $(x,\theta,y)$, which are defined through the following constraint equations to the ``on-shell space'' coordinates $(\lambda,\eta)$~\cite{Huang:2010qy}:
\beqa
\begin{array}{rcrcrlcccc}
x_{i,i+1}^{\alpha \beta} &\equiv&  x_i^{\alpha \beta} &- &x_{i+1}^{\alpha \beta}& =& p_i^{\alpha \beta } &=& \lambda_i^\alpha \lambda_i^\beta 
\\
\theta_{i,i+1}^{I \alpha} &\equiv& \theta_i^{I \alpha} &-& \theta_{i+1}^{I \alpha} &=& q_i^{I\alpha} &=& \lambda_i^\alpha \eta_i^I ,
\\
y_{i,i+1}^{IJ} &\equiv& y_i^{IJ} &-& y_{i+1}^{IJ} &=& r_i^{IJ} &=& \eta_i^I \eta_i^J .
\end{array}
\label{constraint}
\eeqa
where we identify $x_{n+1} \equiv x_1$, $\theta_{n+1}\equiv \theta_1$, $y_{n+1}\equiv y_1$. The dual coordinates are defined such that (super)momentum and part of the R-symmetry are automatically conserved:
\eq
\sum_i p_i=\sum_i q_i=\sum_i r_i=0 \,.
\eqe
Since the constraints eq.(\ref{constraint}) preserve translations in dual space, the dual superconformal invariance of the amplitude can be easily analysed by studying its transformation properties under conformal inversion. The dual space coordinates act on the dual space variables as 
\eq
I[x_i^{\alpha\beta}] =\frac{x_i^{\alpha\beta}}{x_i^2}= -(x^{-1}_i)^{\alpha\beta},~ \; I[\theta_i^{I\alpha}] = \frac{x_i^{\alpha\beta}}{x_i^2} \theta^I_{i\beta} = - (x^{-1}_i)^{\alpha\beta}\theta^I_{i\beta}\;.
\eqe
The spinor indices $(\alpha,\beta)$ are raised and lowered by the antisymmetric $\epsilon$ tensor. The inversion properties of the dual coordinates in combination with eq.(\ref{constraint}) gives the inversion properties of $(\lambda,\eta)$~\cite{Huang:2010qy}:
\eqa
&&I [\lambda_i^\alpha]  = \epsilon_i \frac{(x_i)^{\alpha\beta} \lambda_{i\beta}}{\sqrt{(x_{i+1})^2(x_i)^2}}= \epsilon_i \frac{(x_{i+1})^{\alpha\beta} \lambda_{i\beta}}{\sqrt{(x_{i+1})^2(x_i)^2}},
\;\;\;\;\; (\epsilon_i=\pm 1)  \ \nonumber
\\
&&I [\eta_i^I] = - \epsilon_i \frac{x^2_i}{\sqrt{x_i^2 x_{i+1}^2}} [\eta_i^I +(x^{-1})^{\alpha \beta} \theta_{i \beta}\lambda_{i \alpha}]. 
\eqae
where $\epsilon_i$ is a sign ambiguity that will not affect the final answer. 

The dual conformal symmetry of the tree level amplitude is a statement that under conformal inversion, 
\eq
I\left[\mathcal{A}_n\right]=\prod_{i=1}^n\sqrt{x^2_i}\mathcal{A}_n\,.
\label{Alln}
\eqe
One can rewrite the amplitudes in eq.(\ref{ABJMTree}) utilizing the dual coordinates such that the inversion properties are manifest~\cite{Unpublished}:
\eqa
 \nonumber \mathcal{A}^{tree}_4&=&i\frac{\delta^3(x_1-x_{n+1})\delta^6(\theta_1-\theta_{n+1})}{\sqrt{x^2_{13}x^2_{24}}},\\
 \mathcal{A}^{tree}_6&=&\frac{-i\delta^3(x_1-x_{n+1})\delta^6(\theta_1-\theta_{n+1}) }{x^2_{14}} \left[ \frac{ \left( \frac{\Theta_{413}^I}{\sqrt{x^2_{35}x^2_{13}}}+ i \frac{\Theta_{146}^I}{\sqrt{x^2_{62}x^2_{46}}}  \right)^3}
{(\langle 1\vert  x_{14} \vert 4 \rangle - i \sqrt{x^2_{24}x^2_{51}} )( -\langle 3\vert x_{41} \vert 6 \rangle - i \sqrt{x^2_{13}x^2_{46}})} \right. \nonumber
\\
&&
\left.-\frac{ \left( \frac{\Theta_{413}^I}{\sqrt{x^2_{35}x^2_{13}}}- i \frac{\Theta_{146}^I}{\sqrt{x^2_{62}x^2_{46}}}  \right)^3}
{(\langle 1\vert  x_{14} \vert 4 \rangle +i \sqrt{x^2_{24}x^2_{51}} )( -\langle 3\vert x_{41} \vert 6 \rangle + i \sqrt{x^2_{13}x^2_{46}})}\right]. 
\label{ABJMDCTree}
 \eqae
where $\Theta_{trs}^I\equiv\langle t|x_{tr}x_{rs}\theta_{s}+x_{ts}x_{sr}\theta_{r}+x^2_{sr}\theta_t$ is the dual superconformal covariant function first introduced for the NMHV amplitude of $\mathcal{N}=4$ super Yang-Mills~\cite{DualConformal0}.

Using the fact the recursion relations suitable for ABJM preserves this inversion property, one deduces eq.(\ref{Alln}) from the fact that the four-point amplitude inverts in the desired form~\cite{Gang:2010gy}. At loop level, eq.(\ref{Alln}) is a statement at the integrand level, i.e. prior to regularization. If one factors out the tree level amplitude, which can easily be done for the four-point amplitude, then the remaining integrand $I_n$ inverts as 
\eq
I[I_n]=I_n\,.
\eqe 
We will use this property to construct a basis of integrals.

\subsection{Five-Dimensional Notations}
While dual conformal properties of Lorentz dot products can be conveniently analysed using conformal inversion, it becomes more complicated for tensorial objects such as contractions with Levi-Civita tensors. For these tensorial objects it is simpler to utilize the fact that the conformal symmetry in three dimensions is the same as the Lorentz invariance in five dimensions with signature $(-,-,+,+,+)$. Identifying the three-dimensional Minkowski space as the projective light-cone in five dimensions, conformal invariant objects can be rewritten as Lorentz invariants in five-dimensional notations. 

Consider the homogeneous coordinates in five dimensions as ($T,U,V,W,Y$), the light-cone is given by 
\eq
-T^2-U^2+V^2+W^2+Y^2=0.
\eqe
The light-cone condition removes one degree of freedom. Furthermore, the light-cone condition is invariant under rescaling  ($T\rightarrow \rho T,U\rightarrow \rho U,V\rightarrow \rho V,W\rightarrow \rho W,Y\rightarrow \rho Y$). Identifying points under this rescaling gives us $5-2=3$ degrees of freedom, which is the degree of freedom for a three-dimensional space. Since points in three-dimensional Minkowski space is a null vector in five dimensions, the difference of two points in three dimensions becomes an inner dot product:
\eq
(x_i-x_j)^2=(X_i-X_j)^2=-2X_i\cdot X_j.
\eqe

\begin{figure}[h]
\centering
\includegraphics[scale=0.5]{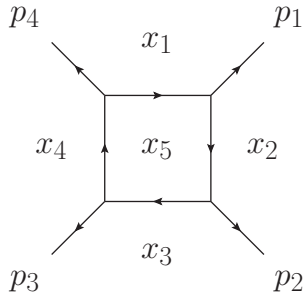}
\caption{Dual conformal coordinates for one-loop diagram}
\label{4ptfigure}
\end{figure}

With a simple exercise, one can convince oneself that there are no dual conformal integrals with scalar numerators at four-point. However, using five-dimensional notations, one can easily identify the following dual conformal integral with tensorial numerator:\footnote{This integrand was first shown to us by Simon Caron-Hout.}
\eq
I^{1-loop}_4=\int \mathcal{D}^3X_5\frac{4\epsilon(5,1,2,3,4)
}{X_{51}^2 X_{52}^2 X_{53}^2  X_{54}^2}.
\label{integrand1}
\eqe
The positions of the dual space coordinates are shown in fig.\ref{4ptfigure}. The measure $\mathcal{D}^3X_5$ is defined as a top form on $\mathbf{RP}^4$,

 \eq
\mathcal{D}^3X=\oint_{S_1} \frac{X\wedge dX\wedge dX\wedge dX\wedge dX}{X^2}.
\eqe
The contour circles the simple pole at $X^2=0$, hence restricts the $\mathbf{RP}^4$ coordinates to the projective light-cone, i.e. the integration is only over the conformally compactified three-dimensional Minkowski space. Conformal invariance comes from the fact that it is 5D Lorentz invariant. Furthermore one can check that the above integral is invariant if one scales any one of the coordinates, which is necessary for it to be a projective integral.\footnote{This approach was first used on integrands in four dimensions, where the embedding space is six dimensions. For explicit applications see \cite{Mason:2010pg,CaronHuot:2011ky}.} 
\section{One-Loop Four-Point Amplitude (an Exercise) \label{OneLoop}}
 In this section, we first study the four-point one-loop amplitude. While the amplitude is expected to vanish, the naive evaluation of the two-particle cut is non-vanishing. We will show that there exists a non-vanishing integrand that satisfies the unitarity cuts, and after integration it does vanish. 

\subsection{The One-Loop Four-Point Integrand}
We begin with the integrand proposed in eq.(\ref{integrand1}). To match with unitarity cuts, we first translate this object back into three-dimensional coordinates. Using light-cone coordinates, the five vectors can be parameterized as 
\eq
i=1-5, X_i=\left(\begin{array}{c} \frac{1}{2}x^2_i \\1  \\\vec{x}_i\end{array}\right),
\eqe
where now $\vec{x}_i$ is a three-dimensional vector. The term $\epsilon(5,1,2,3,4)$ is essentially a determinant 
\eq
\epsilon(5,1,2,3,4)=\left|\begin{array}{ccccc}\frac{1}{2}x^2_5 & \frac{1}{2}x^2_1 & \frac{1}{2}x^2_2& \frac{1}{2}x^2_3 & \frac{1}{2}x^2_4 \\ 1 & 1 & 1 & 1 & 1 \\ \vec{x}_5 & \vec{x}_1 & \vec{x}_2 & \vec{x}_3 &  \vec{x}_4\end{array}\right| .
\eqe
We can choose the origin of the three-dimensional dual space to be at $x_1$, i.e. $\vec{x}_1=0$, one then has 
\eq
\epsilon(5,1,2,3,4)=\left|\begin{array}{ccccc}\frac{1}{2}x^2_5 & 0 & \frac{1}{2}x^2_2& \frac{1}{2}x^2_3 & \frac{1}{2}x^2_4 \\ 0 & 1 & 0 & 0 & 0 \\ \vec{x}_5 & 0 & \vec{x}_2 & \vec{x}_3 &  \vec{x}_4\end{array}\right| .
\eqe
This is not quite correct since the $x_i$s have extra constraints among themselves due to the masslessness of the external momenta. For example, \eq
p^2_1=0\rightarrow (x_1-x_2)^2=0.
\eqe
For our choice of $\vec{x}_1$ this implies $x^2_2=0$. Similarly one has, 
 \eq
p^2_4=(x_4-x_1)^2=0\rightarrow x_4^2=0.
\eqe
Thus we we now have, 
\eqa
\nonumber \epsilon(5,1,2,3,4)&=&\left|\begin{array}{ccccc}\frac{1}{2}x^2_5 & 0 & 0& \frac{1}{2}x^2_3 & 0 \\ 0 & 1 & 0 & 0 & 0 \\ \vec{x}_5 & 0 & \vec{x}_2 & \vec{x}_3 &  \vec{x}_4\end{array}\right|\\
\nonumber&=&\frac{1}{2}\left(x_5^2\epsilon_{\mu\nu\rho}x_2^\mu x_3^\nu x_4^\rho+x_3^2\epsilon_{\mu\nu\rho}x_5^\mu x_2^\nu x_4^\rho\right)\\
&=&\frac{1}{2}\left(x_{51}^2\epsilon_{\mu\nu\rho}x_{21}^\mu x_{31}^\nu x_{41}^\rho+x_{31}^2\epsilon_{\mu\nu\rho}x_{51}^\mu x_{21}^\nu x_{41}^\rho\right),
\eqae
where in the last line we restored $\vec{x}_1$ which was taken to be the origin by in general can be any point. Now the integral in eq.(\ref{integrand1}) can be written in three-dimensional notation:
\eq
I^{1-loop}_4=\int \frac{d^3x_5}{(2\pi)^3}\frac{2x_{51}^2\epsilon_{\mu\nu\rho}x_{21}^\mu x_{31}^\nu x_{41}^\rho+2x_{31}^2\epsilon_{\mu\nu\rho}x_{51}^\mu x_{21}^\nu x_{41}^\rho}{x^2_{15}x^2_{25}x^2_{35}x^2_{45}}\,.
\label{integrand2}
\eqe
One can see that the three-dimensional form obscures its dual conformal properties that were apparent in five dimensions. For the purpose of matching unitarity cuts we further rewrite this in terms of momenta 
\eqa
 I^{1-loop}_4&=&\int \frac{d^3l_1}{(2\pi)^3}\frac{2l_{1}^2\epsilon_{\mu\nu\rho}p_{1}^\mu p_2^\nu p_{4}^\rho+2s\epsilon_{\mu\nu\rho}l_{1}^\mu p_{1}^\nu p_{4}^\rho}{l_1^2(l_1-p_1)^2(l_1-p_1-p_2)^2(l_1+p_4)^2}.
\label{integrand3}
\eqae
We can now analyse the above integrand on the $s$- and $t$-channel cuts.

\subsection{One-Loop Amplitude through Unitarity Cuts}
\begin{figure}
\centering \includegraphics[scale=0.5]{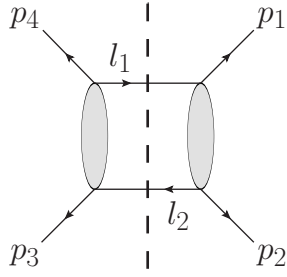}
  \caption{The $s$-channel cut of the one-loop four-point amplitude}\label{4ptcut}
  \end{figure}
For one-loop one may cut at most three propagators in three dimensions. For a four-point amplitude this will lead to three-point tree amplitudes which vanishes on-shell even with complex momenta. Therefore here we consider instead the two-particle cuts. This is illustrated in fig.\ref{4ptcut}.
The s-channel cut is computed by:
\eqa
C_{s}=\delta^3(P)\int d^3\eta_{l_1}d^3\eta_{l_2}\frac{-i\delta^6(Q_L)}{\langle21\rangle\langle 1l_1\rangle}\frac{i\delta^6(Q_R)}{\langle l_2l_1\rangle\langle  l_14\rangle},
\eqae
where we use the convection $\lambda^\alpha_{-l_k}=i\lambda^\alpha_{l_k}$, and 
\eqa
\begin{array}{rcl}
\displaystyle\delta^6(Q_R)&=&\displaystyle\prod^3_{I=1}\delta( q^{I\alpha}_1+q^{I\alpha}_2-q^{I\alpha}_{l_1}+q^{I\alpha}_{l_2})\delta(q^I_{1\alpha}+q^I_{2\alpha}-q^I_{l_1\alpha}+q^I_{l_2\alpha}),\\
\displaystyle\delta^6(Q_L)&=&\displaystyle\prod^3_{I=1}\delta( q^{I\alpha}_3+q^{I\alpha}_4+q^{I\alpha}_{l_1}-q^{I\alpha}_{l_2})\delta(q^I_{3\alpha}+q^I_{4\alpha}+q^I_{l_1\alpha}-q^I_{l_2\alpha}).
\end{array}
\eqae
Combining the delta functions and a straightforward integration gives:
\eqa
\nonumber\int d^3\eta_{l_1}d^3\eta_{l_2}\delta^6(Q_R)\delta^6(Q_L)=\delta^6(Q_{full})\int d^3\eta_{l_1}d^3\eta_{l_2}\delta^6(Q_L)=-\delta^6(Q_{full})\langle l_1l_2\rangle^3,
\eqae
and hence
\eqa
 C_{s}=\delta^3(P)\delta^6(Q_{full})\frac{\langle l_1l_2\rangle^3}{\langle21\rangle\langle 1l_1\rangle\langle l_2l_1\rangle\langle  l_14\rangle}
=-i \mathcal{A}^{tree}_4\frac{\langle12\rangle^2\langle l_11\rangle\langle14\rangle\langle 4l_1\rangle}{(l_1+ p_4)^2 (l_1- p_1)^2}.
\label{cut}
\eqae

We can now compare the $s$-cut of eq.(\ref{integrand3}) with the above eq.(\ref{cut}). The $s$-cut of $I_4^{1-loop}$ is given as 
\eqa
 I^{1-loop}_4|_{s-cut}=\frac{2s \epsilon_{\mu\nu\rho}l_{1}^\mu p_{1}^\nu p_{4}^\rho}{(l_1-p_1)^2(l_1+p_4)^2}
=-\frac{\langle 12\rangle^2\langle l_11\rangle\langle 14\rangle\langle 4l_1\rangle}{(l_1+p_4)^2(l_1-p_1)^2}\,,
\label{Iscut}
\eqae
where we have used the identity 
 \eq
\epsilon_{\mu\nu\rho}=\frac{1}{2}Tr(\si_\mu\si_\nu\si_\rho)\,.
\eqe
Thus comparing eq.(\ref{Iscut}) with eq.(\ref{cut}) one can conclude that 
\eqa
\mathcal{A}_4^{1-loop}&=&i \mathcal{A}^{tree}_4 I_4^{1-loop}.
\eqae
A similar calculation will show that the above result matches with the $t$-channel cut as well.

\subsection{Vanishing of the One-Loop Amplitude\label{vanish1}}
Now that we have the four-point one-loop integrand, it can be straightforwardly integrated. It turns out that the five-dimensional notation is easiest to work with. We begin by using Feynman parametrization to rewriting $I_4^{1-loop}$ in eq.(\ref{integrand1})
\eq
I^{1-loop}_4=2\epsilon_{\mu\nu\rho\sigma\tau} X_1^\nu X_2^\rho X_3^\sigma X_4^\tau \int^1
_0 dF \frac{\partial}{\partial Y^\mu(\alpha)}\int \mathcal{D}^3 X_5 \left[-2X_5\cdot Y(\alpha)\right]^{-3},
\eqe
where $Y(\alpha)=\sum_{i=1}^4\alpha_iX_i$ and $dF=\prod_{i=1}^4 d\alpha_i\delta(1-\sum^4_{i=1}\alpha_i)$. The explicit integration can be done easily, for example going back to three-dimensional notations. One obtains 
\eq
I^{1-loop}_4=\frac{\Gamma(-\frac{3}{2}+3)}{(4\pi)^{3/2}}\epsilon_{\mu\nu\rho\sigma\tau} X_1^\nu X_2^\rho X_3^\sigma X_4^\tau\int^1_0 dF \frac{\partial}{\partial Y^\mu(\alpha)}\left[-Y(\alpha)\cdot Y(\alpha)\right]^{3/2-3}.
\eqe
One can easily see the integrand is proportional to $\epsilon(Y(\alpha)1234)$. Since $Y(\alpha)$ is a linear combination of $X_i$s, $\epsilon(Y(\alpha)1234)$ vanishes. 

Note that this computation is valid only when the dimensions are strictly three. Since there are potential infrared divergences, applying dimensional reduction will potentially invalidate the above argument. As we show in appendix \ref{vanish}, explicit computation using dimensional reduction indeed gives vanishing result up to $\mathcal{O}(\epsilon)$.
\section{Two-Loop Integrand\label{TwoLoop}}
\begin{figure}
\includegraphics[scale=1]{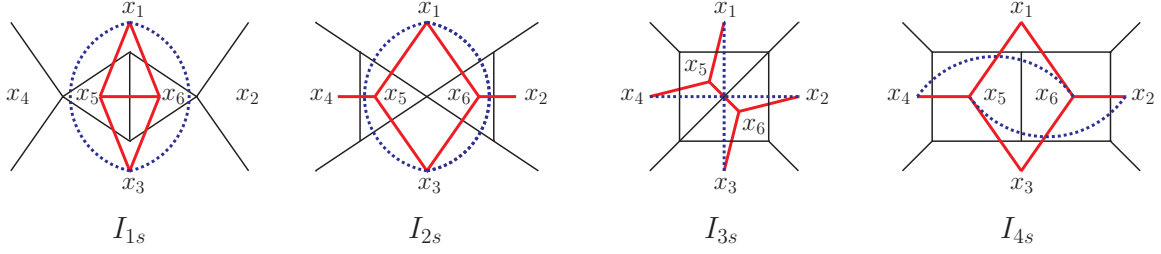}
\caption{The dual conformal invariant integrands with scalar numerators. For each integrand, the (red) solid lines represent propagators while the (blue) dashed lines stretching between $x_i$ and $x_j$ represent the scalar numerator $x_{ij}^2$.}
\label{Basis}
\end{figure}

In this section we construct the two-loop integrand. We begin by constructing a basis of linear independent dual conformal integrals with relative coefficients fixed by requiring the integrand to reproduce the correct $s$-channel double-cut and the three-particle cut. A list of dual conformal integrals at two-loop were already given in~\cite{Huang:2010qy}:
 \beqa
  I_{1s}&\equiv&\int \frac{d^3x_5d^3x_6}{(2\pi)^6}\frac{x^4_{13}}{x^2_{51}x^2_{53}x^2_{56}x^2_{61}x^2_{63}},\;\; \hspace{2cm}I_{1t}=I_{1s}|_{s\rightarrow t}, \\
  I_{2s}&\equiv&\int \frac{d^3x_5d^3x_6}{(2\pi)^6}\frac{x^4_{13}x^2_{42}}{x^2_{51}x^2_{53}x^2_{54}x^2_{61}x^2_{62}x^2_{63}},\;\;\hspace{1.47cm} I_{2t}=I_{2s}|_{s\rightarrow t},\\
  I_{3s}&\equiv&\int  \frac{d^3x_5d^3x_6}{(2\pi)^6}\frac{x^2_{13}x^2_{42}}{x^2_{51}x^2_{54}x^2_{56}x^2_{62} x^2_{63}},\;\;\hspace{2cm} I_{3t}=I_{3s}|_{s\rightarrow t},\\
  I_{4s}&\equiv&\int  \frac{d^3x_5d^3x_6}{(2\pi)^6}\frac{x^4_{13}x^2_{52}x^2_{64}}{x^2_{51}x^2_{53}x^2_{54}x^2_{56}x^2_{61}x^2_{62}x^2_{63}},\;\; \hspace{0.92cm} I_{4t}=I_{4s}|_{s\rightarrow t}.
  \eeqa
where $s\rightarrow t$ corresponds to the inclusion of integrals that are cyclic rotations ($1\rightarrow4, 2\rightarrow1$ and so on) of the original ones. These integrals have simple Lorentz scalar inner products as their numerators, and their graphical representations are shown in fig.\ref{Basis}. 

In light of the role that integrals with Levi-Civita tensors  played for the one-loop integrand, we also include the corresponding contribution at two-loop 

\beqa
\nonumber I_{0s}&\equiv& \int \mathcal{D}^3X_5\mathcal{D}^3X_6 \frac{16\epsilon(5,1,2,3,4)\epsilon(6,1,2,3,4)}{X_{51}^2X_{53}^2X_{54}^2X_{56}^2X_{61}^2X_{63}^2X_{62}^2X_{42}^2}\\
&=& \int \frac{d^3x_5d^3x_6}{(2\pi)^6} \frac{\xi_s
}{x_{51}^2 x_{53}^2 x_{54}^2x_{56}^2 x_{61}^2 x_{62}^2 x_{63}^2 x_{24}^2},
\label{tensor}
\eeqa
where $\xi_s\equiv 4\epsilon_{\mu\nu\rho}(x_{51}^2x_{21}^\mu x_{31}^\nu x_{41}^\rho+x_{31}^2 x_{51}^\mu x_{21}^\nu x_{41}^\rho)\epsilon_{\gamma\sigma\eta}
(x_{61}^2x_{21}^\gamma x_{31}^\sigma x_{41}^\eta+x_{31}^2 x_{61}^\gamma x_{21}^\sigma x_{41}^\eta)$.
Not surprisingly, these five integrals are not linearly independent. To see this, one can convert the product of Levi-Civita tensors into Lorentz inner products by using the following identity, 
\beqa
\label{epdel}\epsilon_{\mu\nu\rho}\epsilon^{\eta\tau\la}=-\de_{[\mu\nu\rho]}^{\eta\tau\la}.
\eeqa
From this one can show that:
\beqa
2I_{0s}=I_{1s}-I_{2s}+I_{3s}+I_{3t}+I_{4s}.
\label{I0sIdentity}
\eeqa
Thus we see that one can trade the double-box integral $I_{4s}$ in terms of the other double-box integral $I_{0s}$. In the following analysis, we choose $I_{0s}$ as our only double-box integral.

\begin{figure}[h]
\centering
 \subfigure[Double-$s$-channel cut]{\includegraphics[scale=1.4]{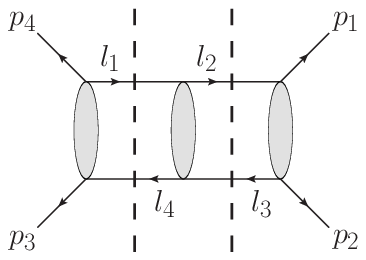}}
  ~~~~~~~~~~~~~~~~~~~~~
  \subfigure[Three-particle cut]{\includegraphics[scale=1.4]{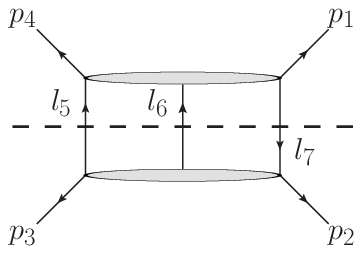}}
  \caption{(a) Double-$s$-channel cut diagram shows the two-loop diagram can be form by sewing three four-point tree diagrams together. (b) Three-particle cut shows the two-loop diagram can be stuck with two five-point tree diagrams which should vanish individually.}
  \label{2loopst}
  \end{figure}

We thus begin with the following ansatz for the four-point two-loop amplitude:
\beqa
\label{ANS}\mathcal{A}_4^{2-loop}=\mathcal{A}_4^{tree}\sum_{i=0}^3\left[c_{is}I_{is}+\left(s\rightarrow t\right)\right],
\eeqa
where $\mathcal{A}_4^{tree}=i\de^3(P)\delta^6(Q_{full})/\langle41\rangle\langle12\rangle$. The coefficients eq.(\ref{ANS}) will be fixed by matching with the double-$s$-channel cut, 
and the three-particle cut as shown in fig.\ref{2loopst}.
\subsection{Double-$s$-Channel Cut}
The double-$s$-channel cut, as shown in fig.\ref{2loopst}.a,  is computed by sewing together three four-point tree amplitudes,
\beqa
C_s=\de^3(P)\int d^3\eta_{l_1}d^3\eta_{l_2}d^3\eta_{l_3}d^3\eta_{l_4}\frac{i\delta^6(Q_1)}{\langle 21\rangle\langle 1 -l_2\rangle}\frac{i\delta^6(Q_2)}{\langle -l_3 l_2\rangle\langle l_2 -l_1\rangle}\frac{i\delta^6(Q_3)}{\braket{-l_4l_1}\langle l_14 \rangle
}.
\eeqa
The delta functions $\de^6(Q_k)$s are given by,
\beqa
\begin{array}{rclll}
\displaystyle\delta^6(Q_1)&=&\displaystyle\prod^3_{I=1}&\delta( q^{I\alpha}_3+q^{I\alpha}_4+q^{I\alpha}_{l_1}-q^{I\alpha}_{l_4})
&\delta( q^{I}_{3\al}+q^{I}_{4\al}+q^{I}_{l_1\al}-q^{I}_{l_4\al}),\\
\displaystyle\delta^6(Q_2)&=&\displaystyle\prod^3_{I=1}&\delta(q^{I\alpha}_{l_1}-q^{I\alpha}_{l_2}+q^{I\alpha}_{l_3}-q^{I\alpha}_{l_4})&\delta( q^{I}_{l_1\al}-q^{I}_{l_2\al}+q^{I}_{l_3\al}-q^{I}_{l_4\al}),\\
\displaystyle\delta^6(Q_3)&=&\displaystyle\prod^3_{I=1}&\delta( q^{I\alpha}_1+q^{I\alpha}_2-q^{I\alpha}_{l_2}+q^{I\alpha}_{l_3})&\delta( q^{I}_{1\al}+q^{I}_{2\al}-q^{I}_{l_2\al}+q^{I}_{l_3\al}).
\end{array}
\eeqa
The result from integrating over the $\eta$s give:
\beqa
C_s=-i\frac{\delta^3(P)\delta^6(Q_{full})\langle l_1l_4\rangle^3\langle l_2l_3\rangle^3}{\langle 21\rangle\langle 1 l_2\rangle\langle l_3 l_2\rangle\langle l_2 l_1\rangle\braket{l_4l_1}\langle l_14 \rangle}
=\mathcal{A}_4^{tree} \varrho_1 s^2 \braket{14}\braket{4l_1}\braket{l_1 l_2}\braket{l_2 1}\,,
\label{spcut}
\eeqa
where $\rho_1^{-1}=\langle 4l_1\rangle^2\langle l_1l_2\rangle^2\langle l_21\rangle^2$.

Only integrals $I_{0s}$ ,$I_{1s}$ ,$I_{2s}$ contribute to the double-$s$-channel cut. Their contributions are, respectively,

   \beqa
   \begin{array}{rcl}
   I_{0s}|_{s-cut}
&=&\displaystyle \left. \frac{4x_{31}^4
\epsilon_{\mu\nu\rho}\epsilon_{\gamma\sigma\eta}
x_{51}^\mu x_{21}^\nu x_{41}^\rho
 x_{61}^\gamma x_{21}^\sigma  x_{41}^\eta}{ x_{54}^2x_{56}^2  x_{62}^2  x_{24}^2}\right|_{l_s^2=0}\\
&=&s^2\varrho_1
\big(\braket{l_14}^2\braket{l_2 1}^2
-\braket{14}\braket{4l_1}\braket{l_1l_2}\braket{l_21}\big),\\
  I_{1s}|_{s-cut}
&=&\displaystyle\left.\frac{x^4_{13}}{x^2_{56}}\right|_{l_s^2=0}\hspace{0.53cm}= s^2\varrho_1
\braket{l_14}^2 \braket{l_21}^2,\\
 I_{2s}|_{s-cut}
&=&\displaystyle\left.\frac{x^4_{13}x^2_{42}}{x^2_{54}x^2_{62}}\right|_{l_s^2=0}= s^2\varrho_1\braket{14}^2 \braket{l_1l_2}^2,\\
\end{array}
 \eeqa
where $|_{l_s^2=0}$ indicates the on-shell conditions on the double-$s$-channel cut propagators, $l^2_1=l^2_2=l^2_3=l^2_4=0$, and Schouten identities have been applied to put this result into spinor inner products that are linearly independent, i.e. further applications of Schouten identities will generate new spinor inner products. Matching the double-$s$-channel cut implies solving 
\beqa
 \big( c_{0s}I_{0s}+c_{1s}I_{1s}+c_{2s}I_{2s}\big)|_{s-cut}=\varrho_1 s^2 \braket{14}\braket{4l_1}\braket{l_1 l_2}\braket{l_2 1},
\label{scut}
\eeqa
for some coefficient $c_{0s}$,  $c_{1s}$, $c_{2s}$. Since the contribution from various integrals are now in independent basis, one can deduce,
\beqa
\label{cs} c_{0s}=-1,~
c_{2s}=0,~c_{1s}=1.
\eeqa
Note that the double-$s$-channel cut condition will not give any information about the coefficient of $I_{3s}$. This integral will contribute to the $t$-channel three-particle cut, which we will evaluate in the next subsection.
\subsection{Three-Particle Cut}
\begin{figure}[h]\centering
\subfigure[$I_{0t}$]{\includegraphics[scale=0.6]{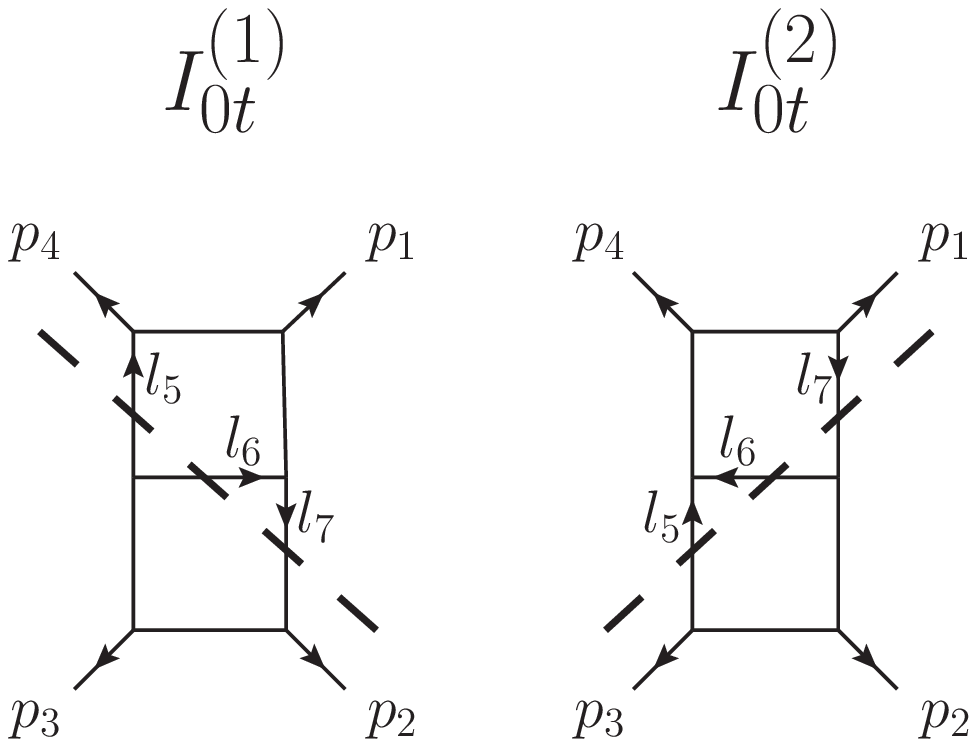}}~~~~~~~~~~~~~~~~~~~~~~
\subfigure[$I_{1t}$]{\includegraphics[scale=0.6]{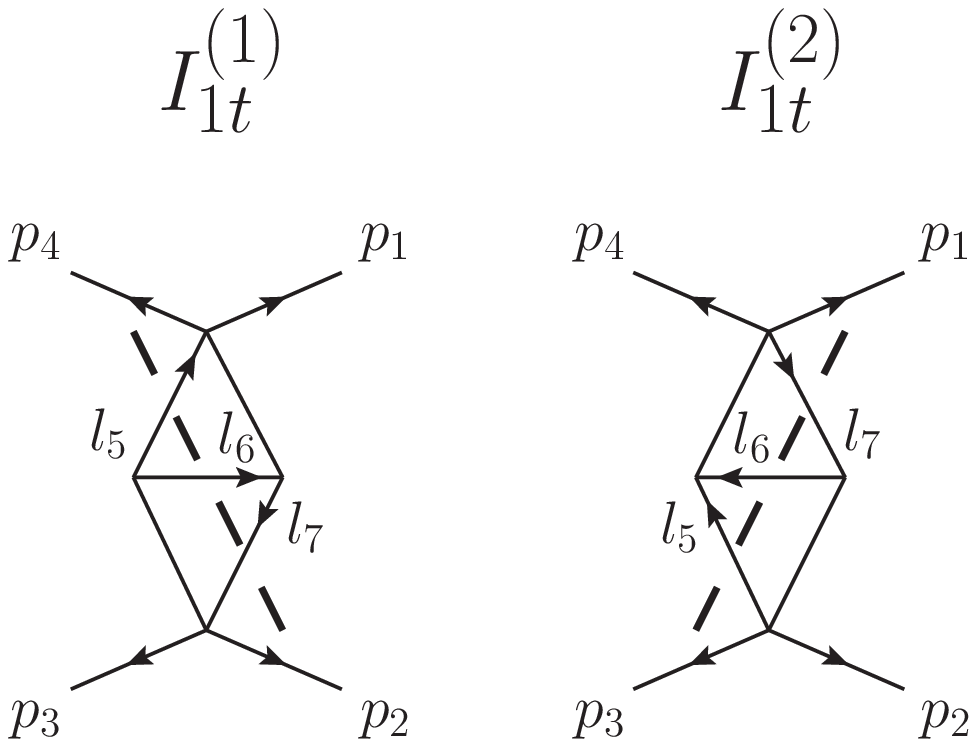}}
  \caption{Here shows that for each single diagram $I_{0t}$ or $I_{1t}$, one can have two different kinds of three-particle cut.}
  \label{superindex}
\end{figure}

The three-particle cut shown in fig.\ref{2loopst}.b will simply be zero due to the fact that odd tree-level amplitudes vanish. Note that the cyclic rotated integrals will contribute to this cut as well. Here we list all of them,
  \beqa
  I_{0t}&\equiv&\int \frac{d^3x_5d^3x_6}{(2\pi)^6}\frac{\xi_t
}{x_{51}^2 x_{52}^2 x_{54}^2 x_{56}^2 x_{62}^2 x_{63}^2 x_{64}^2 x_{13}^2},\\
  I_{1t}&\equiv&\int \frac{d^3x_5d^3x_6}{(2\pi)^6}\frac{x^4_{24}}{x^2_{52}x^2_{54}x^2_{56}x^2_{62}x^2_{64}},\\
  I_{2t}&\equiv&\int \frac{d^3x_5d^3x_6}{(2\pi)^6}\frac{x^4_{24}x^2_{13}}{x^2_{51}x^2_{52}x^2_{54}x^2_{62}x^2_{63}x^2_{64}},\\
   I_{3t}&\equiv&  
  \int \frac{d^3x_5d^3x_6}{(2\pi)^6}\frac{x^2_{13}x^2_{42}}{x^2_{53}x^2_{54}x^2_{56}x^2_{62}x^2_{61}},
\eeqa
  where $\xi_t\equiv \xi_s|_{s\leftrightarrow t}=4\epsilon_{\mu\nu\rho}(x_{52}^2x_{32}^\mu x_{42}^\nu x_{12}^\rho+x_{42}^2 x_{52}^\mu x_{32}^\nu x_{12}^\rho)\epsilon_{\gamma\sigma\eta}
(x_{62}^2x_{32}^\gamma x_{42}^\sigma x_{12}^\eta+x_{42}^2 x_{62}^\gamma x_{32}^\sigma x_{12}^\eta)$.
The integrals, contributing to the three-particle cut are $I_{0s}$ ,$I_{3s}$ ,$I_{0t}$ ,$I_{1t}$ ,$I_{3t}$, and their contributions are,
\beqa
\begin{array}{rcl}
 I_{0s}|_{3p-cut}
&=&\displaystyle\left.\frac{\xi_s
}{x_{51}^2 x_{53}^2  x_{61}^2 x_{63}^2 x_{24}^2}\right|_{l_t^2=0}=-\varrho_2 \braket{12}^2\braket{14}^2\braket{l_53}\braket{l_54}\braket{l_71}\braket{l_72}
\braket{l_5l_6}^2\braket{l_6l_7}^2,\\
 I_{3s}|_{3p-cut}
&=&\displaystyle\left.\frac{x^2_{13}x^2_{24}}{x^2_{53}x^2_{61}}\right|_{l_t^2=0}
=\varrho_2\braket{12}^2\braket{14}^2\braket{l_54}^2\braket{l_72}^2\braket{l_5l_6}^2\braket{l_6l_7}^2,\\
 I_{0t}^{(1)}|_{3p-cut}&=&\displaystyle\left.\frac{\xi_t
}{x_{13}^2 x_{51}^2 x_{52}^2  x_{63}^2 x_{64}^2 }\right|_{l_t^2=0}=\varrho_2\braket{14}^4\braket{l_53}^3\braket{l_54}\braket{l_71}^3\braket{l_72},\\
 I_{0t}^{(2)}|_{3p-cut}&=&\displaystyle\left.\frac{\xi_t
}{x_{13}^2 x_{51}^2 x_{54}^2 x_{62}^2 x_{63}^2  }\right|_{l_t^2=0}
=\varrho_2\braket{14}^4\braket{l_53}\braket{l_54}^3\braket{l_71}\braket{l_72}^3,\\
I_{1t}^{(1)}|_{3p-cut}&=&\displaystyle\left.\frac{x^4_{24}}{x^2_{52}x^2_{64}}\right|_{l_t^2=0}
=-\varrho_2 
\braket{14}^4\braket{l_54}^2\braket{l_53}^2\braket{l_71}^2\braket{l_72}^2,\\
I_{1t}^{(2)}|_{3p-cut}&=&I_{1t}^{(1)}|_{3p-cut},\\
 I_{3t}|_{3p-cut}&=&\displaystyle\left.\frac{x^2_{13}x^2_{24}}{x^2_{51}x^2_{63}}\right|_{l_t^2=0}
=\varrho_2 
\braket{12}^2\braket{14}^2\braket{l_53}^2\braket{l_71}^2\braket{l_5l_6}^2\braket{l_6l_7}^2.
\end{array}
\eeqa
where 
$\varrho_2^{-1}\equiv\braket{l_53}^2\braket{l_54}^2\braket{l_71}^2\braket{l_72}^2\braket{l_5l_6}^2\braket{l_6l_7}^2$
and,  in $I_{0s}$ and $I_{0t}$, the superindices for $I_{0t},I_{1t}$ indicate the two contributions from the same diagram as in Fig.\ref{superindex}.

We convert all spinor inner products into an independent basis consisting of $\om_1\equiv\braket{l_53}\braket{l_71}$ and $\om_2\equiv\braket{l_54}\braket{l_72}$ along with spinor brackets involving only external momenta. This can be achieved by using the identity,
\beqa
\label{ttt}\braket{12}\braket{l_5l_6}\braket{l_6l_7}
&=&\braket{41}(\om_1+\om_2).
\eeqa
One can prove this identity by using momentum conservation $l_6=-l_5+l_7+p_1+p_4$, and Schouten identities.
With the helps of eq.(\ref{ttt}), we can express all integrals as\footnote{Note that we do not convert $\varrho_2$ into $(\om_1,\om_2)$ since it appears as an overall factor for all integrals.} 
\beqa
\begin{array}{rcl}
 I_{0s}|_{3p-cut}
&=&-t^2\varrho_2\big(\om_1^3\om_2+2\om_1^2\om_2^2+\om_1 \om_2^3\big),\\
 I_{3s}|_{3p-cut}&=&t^2\varrho_2\big(\om_1^4+2\om_1^3\om_2+\om_1^2\om_2^2\big),\\
 I_{0t}^{(1)}|_{3p-cut}&=&t^2\varrho_2\om_1^3\om_2,\\
 I_{0t}^{(2)}|_{3p-cut}&=&t^2\varrho_2\om_1\om_2^3,\\
 I_{1t}^{(1)}|_{3p-cut}&=& -t^2\varrho_2 \om_1^2\om_2^2,\\
 I_{1t}^{(2)}|_{3p-cut}&=& I_{1t}^{(1)}|_{3p-cut},\\ 
 I_{3t}|_{3p-cut}&=&t^2\varrho_2\big(\om_1^2\om_2^2+2\om_1\om_2^3+\om_2^4\big).
\end{array}
\eeqa
The three-particle cut should be zero as described previously such that each combination in different powers of $\om_1$ and $\om_2$ implies the following constraints,
\beqa\label{cef}
\begin{array}{rrl}
\om_1^4:&c_{3s}&=0,\\
\om_1^3\om_2:&-c_{0s}+2c_{3s}+c_{0t}&=0,\\
\om_1^2\om_2^2:&-2c_{0s}+c_{3s}-2c_{1t}+c_{3t}&=0,\\
\om_1\om_2^3:&-c_{0s}+c_{0t}+2c_{3t}&=0,\\
\om_2^4:&c_{3t}&=0,
\end{array}
\eeqa
By using the already known $c_{is}$s in eq.(\ref{cs}), we can solve part of $c_{it}$s,
\beqa
\label{ct}c_{0t}=-c_{1t}=-1.
\eeqa
One can see that eq.(\ref{ct}) satisfies the double-$t$-channel cut, along with the constraint $c_{2t}=0$. Thus one arrives at,
\beqa
c_{0s}=c_{0t}=-c_{1s}=-c_{1t}=-1,~c_{2s}=c_{2t}=c_{3s}=c_{3t}=0.
\eeqa

Thus in conclusion, the unitarity cuts fix the four-point two-loop integrand of ABJM theory to be:
\eq
\mathcal{A}_4^{2-loop}=\left(\frac{N}{K}\right)^2\mathcal{A}_4^{tree}\left[-I_{0s}+I_{1s}+\left(s\rightarrow t\right)\right].
\label{final}
\eqe
\section{Obtaining the Two-Loop Integral}\label{TL}
In this section we integrate eq.(\ref{final}). Both integrals $I_{0s,t}$ and $I_{1s,t}$ are infrared divergent and require regularization. There are  three different regularization schemes available for Chern-Simons like theories, the usual dimensional regularization, Yang-Mills mass regulator and dimensional reduction. In Yang-Mills mass regulation, one introduces a $\frac{1}{e^2}trF^2$ term to the action. Since the Yang-Mills coupling constant $e$ is dimensionful in three dimensions, it serves as a regulator that is taken to be zero at the end of the calculation. In dimensional reduction, all tensor algebra related to Levi-Civita tensors are performed in strictly three dimensions. Once one obtains scalar integrands, one analytically continues to $D=3-2\epsilon$. Dimensional reduction for Chern-Simons theory has been tested in~\cite{Chen:1992ee} to three-loop order and has been shown to satisfy the Slavnov-Taylor identities, while dimensional regularization fails to produce gauge invariant result at two-loop order. Since dimensional reduction was used for the Wilson loop computation, we will use this as our regularization scheme.

The integrands $I_{1s,t}$ can be straightforwardly integrated using integrating by parts~\cite{Chetyrkin:1981qh} technique. This integral produces only $\mathcal{O}(\epsilon^{-1})$ divergences. The tensor integral $I_{0s,t}$ is more complicated and explicit integration is done using Mellin-Barnes representation technique. This integral yields $\mathcal{O}(\epsilon^{-2})$ divergence.
\subsection{Integrating $I_{1s}$}
\par We now apply the method of integration by parts~\cite{Chetyrkin:1981qh} to the integrand $I_{1s}$, the ``kite" integrand, which in momentum space can be written as, 
\eq
I_{1s}=\int \frac{d^D l_1}{(2\pi)^D}\int \frac{d^D l_2}{(2\pi)^D}\left[\frac{s^2}{l_1^2(l_1-p_1-p_2)^2(l_1-l_2)^2l_2^2(l_2-p_1-p_2)^2}\right].
\eqe 
We begin with the left triangle sub-diagram, where the external lines need not to be massless:
\beqa
\Delta\equiv\int\frac{d^Dl_1}{(2\pi)^D}\frac{1}{l_1^2(l_1-l_2)^2(l_1-M)^2},
\eeqa
where $M\equiv p_1+p_2$. One inserts the following identity operator into the integrand, 
\beqa
1= \frac{1}{D}\frac{\partial}{\partial l^\mu_{1}} (l_1-l_2)^\mu.
\eeqa
Integrating by parts with respect $l_1$ and rewriting the Lorentz inner products in terms of inverse propagators, the $l_1$ part of the integral will generate a vanishing surface term plus four single propagator terms:
\beqa
\label{KITE1}\Delta
=-\frac{1}{D-4}\int \frac{d^D l_1}{(2\pi)^D}\left[\frac{l_2^2}{l_1^2}+\frac{(l_2-M)^2}{(l_1-M)^2}-\frac{(l_1-l_2)^2}{l_1^2}-\frac{(l_1-l_2)^2}{(l_1-M)^2}\right].
\eeqa
Putting back the remaining propagators and use a change of variables for the loop momenta for some of the terms, we can reexpress $I_{1s}$ in terms of two simpler integrals,
\beqa
\nonumber I_{1s}&=&\frac{2s^2}{\left(D-4\right)}\int \frac{d^D l_1}{(2\pi)^D}\int \frac{d^D l_2}{(2\pi)^D}\left[-\frac{1}{l_1^2(l_1-M)^4l_2^2(l_1-l_2)^2}+\frac{1}{l_1^2(l_1-M)^4l_2^2(l_2-M)^2}\right].\\
\eeqa
Explicit integration gives,
\beqa
\notag -I_{1s}&=&\frac{-2s^{D-3}\Ga\left(\frac{D}{2}-1\right)^2\Ga\left(\frac{D}{2}-2\right)\Ga\left(-\frac{D}{2}+2\right)}{\left(4\pi\right)^D\left(D-4\right)\Ga\left(D-2\right)}\\
&&\times\left[\frac{\Ga\left(\frac{D}{2}-1\right)\Ga\left(-\frac{D}{2}+3\right)}{ \Ga\left(D-3\right)}
-\frac{\Ga\left(D-3\right)\Ga\left(-D+5\right)}{\Ga\left(\frac{3D}{2}-5\right)\Ga\left(-\frac{D}{2}+3\right)}
\right].
\eeqa
Inserting $D=3-2\epsilon$ and expanding in $\epsilon$ one obtains:
\eq
\framebox[8.6cm][c]{$\displaystyle{I_{1s}=\frac{-1}{16\pi^2}\left[\frac{1}{2\epsilon}\left(\frac{e^{\gamma_E}s}{8\pi}\right)^{-2\epsilon}+1-\ln 2+\mathcal{O}(\epsilon)\right]}$}~.
\label{KiteResult}
\eqe
\subsection{Integrating $I_{0s}$\label{I0s}}
\par The $I_{0s}$ integral defined in eq.(\ref{tensor}) can be naturally separated into four pieces by expanding the numerator. In the following we will compute each piece individually and give detailed derivation of the term with numerator $s^2\epsilon(l_1p_1p_4)\epsilon(l_2p_1p_4)$ since this gives the most singular $\mathcal{O}(\epsilon^{-2})$ divergence, while the remaining results are listed in the appendix \ref{API0}.   \par

We begin by decomposing $I_{0s}$ into four parts,
\beqa
I_{0s}=I_{01}+I_{02}+I_{03}+I_{04},
\eeqa
where
\beqa
 I_{01}&\equiv&\int \frac{d^D l_1 d^D l_2}{(2\pi)^{2D}}\frac{4s^2t^{-1}\epsilon_{\mu\nu\rho} l_1^\mu p_1^\nu p_4^\rho 
\epsilon_{\si\eta\la} l_2^\si p_1^\eta p_4^\la}{l_1^2(l_1+p_4)^2(l_1+p_3+p_4)^2(l_1-l_2)^2l_2^2 (l_2-p_1)^2(l_2-p_1-p_2)^2},\\
I_{02}&\equiv& \int \frac{d^D l_1 d^D l_2}{(2\pi)^{2D}}\frac{4st^{-1}\epsilon_{\mu\nu\rho} l_1^\mu p_1^\nu p_4^\rho
\epsilon_{\si\eta\la} p_1^\si p_2^\eta p_4^\la
}{l_1^2(l_1+p_4)^2(l_1+p_3+p_4)^2(l_1-l_2)^2(l_2-p_1)^2(l_2-p_1-p_2)^2 },\\
 I_{03}&\equiv&\int \frac{d^D l_1 d^D l_2}{(2\pi)^{2D}}\frac{4st^{-1}\epsilon_{\mu\nu\rho} p_1^\mu p_2^\nu p_4^\rho \epsilon_{\si\eta\la} l_2^\si p_1^\eta p_4^\la }{(l_1+p_4)^2(l_1+p_3+p_4)^2(l_1-l_2)^2 l_2^2 (l_2-p_1)^2(l_2-p_1-p_2)^2},\\
 I_{04}&\equiv&\int \frac{d^D l_1 d^D l_2}{(2\pi)^{2D}}\frac{4 t^{-1}\epsilon_{\mu\nu\rho} p_1^\mu p_2^\nu p_4^\rho\epsilon_{\si\eta\la} p_1^\si p_2^\eta p_4^\la}{(l_1+p_4)^2(l_1+p_3+p_4)^2(l_1-l_2)^2(l_2-p_1)^2(l_2-p_1-p_2)^2}.
\eeqa
As it turns out, the most divergent piece is contained in $I_{01}$, which we now focus on.

We will proceed by first integrating $l_1$, thus we separate the part of the integrand that contains $l_1$ and define
\beqa
I_P&\equiv &\int \frac{d^D l_1}{(2\pi)^D}\frac{\epsilon_{\mu\nu\rho} l_1^\mu p_1^\nu p_4^\rho }{l_1^2(l_1+p_4)^2(l_1+p_3+p_4)^2(l_1-l_2)^2}.
\eeqa
Notice that while the Levi-Civita tensor is three-dimensional, the loop integration measure is taken to be $D$ dimensional. The justification is that as the numerator is linear in $l_1^\mu$, one can perform a change of variables as in eq.(\ref{shift2}) such that the loop-momentum dependent part of the numerator integrates to zero regardless of the dimension. The remaining loop momentum dependence is that of the scalar propagators which one can freely continue to $D=3-2\e$.   Using Feynman parameters and integrating out $l_1$, we can put $I_{P}$ into the form,
\beqa
\label{IP1}\int^1_0 dF \frac{\Ga\left(4-\frac{D}{2}\right)\epsilon_{\mu\nu\rho} (-\al_3 p_3+\al_4 l_2)^\mu p_1^\nu p_4^\rho}{(4\pi)^{\frac{D}{2}}\left[\al_1\al_3 s+\al_1\al_4 l_2^2+\al_2\al_4(l_2+p_4)^2+\al_3\al_4(l_2+p_3+p_4)^2\right]^{4-\frac{D}{2}}}.
\eeqa
To rewrite the denominator as products of propagators, which will be useful for the $l_2$ integration, we convert the above integrand using the Mellin-Barnes(MB) representation:
\beqa
\frac{1}{(X+Y)^v}=\int^{i\infty}_{-i\infty} \frac{dw}{2\pi i}\frac{Y^w}{X^{v+w}}\frac{\Ga(-w)\Ga(v+w)}{\Ga(v)}.
\eeqa
Repeated use of the Mellin-Barnes representation converts eq.(\ref{IP1}) to,
\beqa
\notag &&\int^{+i\infty}_{-i\infty} \frac{dz_1dz_2dz_3}{(2\pi i)^3}\int^1_0 
dF \frac{\epsilon_{\mu\nu\rho} (-\al_3 p_3+\al_4 l_2)^\mu p_1^\nu p_4^\rho }{(4\pi)^{\frac{D}{2}}}\frac{[\al_1\al_4 (l_2)^2]^{z_1}[\al_2\al_4(l_2+p_4)^2]^{z_2}}{(\al_1\al_3 s)^{z_1+z_2+z_3-\frac{D}{2}+4}} \\
\label{AA}&&\times[\al_3\al_4(l_2+p_3+p_4)^2]^{z_3}
\Ga(-z_1)\Ga(-z_2)\Ga(-z_3)\Ga\left(\mbox{$z_1+z_2+z_3-\frac{D}{2}+4$}\right).
\eeqa
The integration of Feynman parameters can be carried out using the formula eq.(\ref{IFP}), and one obtains,
\beqa
\label{A}\framebox[13.1cm][c]{$I_P=Z_1( \epsilon_{\mu\nu\rho}l_2^\mu p_1^\nu p_4^\rho A+ \epsilon_{\mu\nu\rho}p_1p_2p_4 B)[(l_2)^2]^{z_1}[(l_2+p_4)^2]^{z_2}[(l_2-p_1-p_2)^2]^{z_3}$}~,
\eeqa
where
\beqa
\notag 
\notag Z_1&\equiv&\int^{+i\infty}_{-i\infty} \frac{dz_1dz_2dz_3}{(2\pi i)^3}
\frac{s^{-z_1-z_2-z_3+\frac{D}{2}-4}}{(4\pi)^{\frac{D}{2}}\Ga\left(D-3\right)}\Ga(-z_1)\Ga(-z_2)\Ga(-z_3)\\
\label{Z1}&&
\times \Ga\left(z_2+1\right)\Ga\left(\mbox{$z_1+z_2+z_3-\frac{D}{2}+4$}\right)\Ga\left(\mbox{$-z_2-z_3+\frac{D}{2}-3$}\right),\\
 \label{A}A&\equiv& \Ga\left(\mbox{$-z_1-z_2+\frac{D}{2}-3$}\right)\Ga\left(z_1+z_2+z_3+2\right),
\\
 B&\equiv& -\Ga\left(\mbox{$-z_1-z_2+\frac{D}{2}-2$}\right)\Ga\left(z_1+z_2+z_3+1\right).
\eeqa

One can now perform the $l_2$ integration. Inserting the result derived in eq.(\ref{A}) into $I_{01}$, the integrand separates into two pieces:
\beqa
\label{I01} I_{01}=I_{01A}+I_{01B},
\eeqa
where 
\beqa
I_{01A}&\equiv&\int d^D l_2 Z_1\frac{4s^2t^{-1}\epsilon_{\mu\nu\rho}l_2^\mu p_1^\nu p_4^\rho \epsilon_{\si\eta\la} l_2^\si p_1^\eta p_4^\la A}{[(l_2)^2]^{1-z_1} [(l_2+p_4)^2]^{-z_2}[(l_2-p_1-p_2)^2]^{1-z_3}(l_2-p_1)^2},\\
I_{01B}&\equiv&\int d^D l_2 Z_1\frac{4s^2t^{-1} \epsilon_{\mu\nu\rho}p_1^\mu p_2^\nu p_4^\rho \epsilon_{\si\eta\la} l_2^\si p_1^\eta p_4^\la B}{[(l_2)^2]^{1-z_1} [(l_2+p_4)^2]^{-z_2}[(l_2-p_1-p_2)^2]^{1-z_3}(l_2-p_1)^2}.
\eeqa
Here we see again that while the numerators contain three-dimensional Levi-Civita tensors, the integration measure is $D$ dimensional. This is allowed as we are implementing dimensional reduction, where we first use three-dimensional tensor algebra to convert the products of Levi-Civita tensors to products of Kronecker deltas as in eq.(\ref{epdel}). This gives scalar numerators which we can continue to $D$-dimensions. The numerator of $I_{A_1}$ can be rewritten as,
\beqa
\notag &&4\epsilon_{\mu\nu\rho}l_2^\mu p_1^\nu p_4^\rho \epsilon_{\si\eta\la} l_2^\si p_1^\eta p_4^\la\\
&&=t\big[(l_2+p_4)^2(l_2-p_1)^2-l_2^2(l_2-p_1)^2-l_2^2(l_2+p_4)^2+(l_2^2)^2+tl_2^2\big].
\eeqa  
Thus $I_{01A}$ further splits into five terms,
\beqa
\notag I_{01A}=
\label{I11} I_{011}+I_{012}+I_{013}+I_{014}+I_{015},
\eeqa
where
\beqa
I_{011}&\equiv&Z_1\int \frac{d^D l_2}{(2\pi)^D} \frac{As^2t}{[(l_2)^2]^{-z_1} [(l_2+p_4)^2]^{-z_2}[(l_2-p_1-p_2)^2]^{1-z_3}(l_2-p_1)^2},\\
\label{A12}I_{012}&\equiv&Z_1\int \frac{d^D l_2}{(2\pi)^D} \frac{As^2}{[(l_2)^2]^{1-z_1} [(l_2+p_4)^2]^{-1-z_2}[(l_2-p_1-p_2)^2]^{1-z_3}},\\
\label{A13}I_{013}&\equiv&-Z_1\int\frac{d^D l_2}{(2\pi)^D} \frac{As^2}{[(l_2)^2]^{-z_1} [(l_2+p_4)^2]^{-z_2}[(l_2-p_1-p_2)^2]^{1-z_3}},\\
\label{A14}I_{014}&\equiv&-Z_1\int \frac{d^D l_2}{(2\pi)^D} \frac{As^2}{[(l_2)^2]^{-z_1} [(l_2+p_4)^2]^{-1-z_2}[(l_2-p_1-p_2)^2]^{1-z_3}(l_2-p_1)^2},\\
\label{A15}I_{015}&\equiv&Z_1\int \frac{d^D l_2}{(2\pi)^D} \frac{As^2}{[(l_2)^2]^{-1-z_1} [(l_2+p_4)^2]^{-z_2}[(l_2-p_1-p_2)^2]^{1-z_3}(l_2-p_1)^2}.
\eeqa
We complete the $l_2$ integral for $I_{011}$ as an example, since it is the only $s$-channel integral that contains the $\mathcal{O}(\epsilon^{-2})$ divergence. Again using Feynman parameterization and an additional MB representation, we obtain,  
\beqa
\notag \textstyle I_{011} &=&\int^{+i\infty}_{-i\infty} 
\frac{dz_1dz_2dz_3dz_4}{(4\pi)^D (2\pi i)^4}
\frac{ s^{-z_4+D-4}t^{z_4+1}\Ga(z_1+z_2+z_3-\frac{D}{2}+4)\Ga(z_1+z_2+z_3+2)
}{\Ga(D-3)\Ga(-z_3+1)\Ga(z_1+z_2+z_3+D-2)}\\
 \notag&&\textstyle\times\Ga(z_1+z_2-z_4+\frac{D}{2}-1) \Ga(z_2+z_3-z_4+\frac{D}{2}-2)\Ga(z_2+1)\Ga(z_4+1)\\
  \notag&&\textstyle\times\Ga(-z_1-z_2-z_3+z_4-\frac{D}{2}+2)\Ga(-z_2-z_3+\frac{D}{2}-3)\Ga(-z_1-z_2+\frac{D}{2}-3)\\
&&\times\Ga(-z_2+z_4)\Ga(-z_3)\Ga(-z_4).
\eeqa
Explicit integration can be carried out using the $Mathemtica$ package MB.m~\cite{Czakon:2005rk}. The above integral $I_{011}$ plus the its $t$-channel counter part gives:
\eqa
\nonumber I_{011}+(s\leftrightarrow t)
\notag =\frac{1}{16\pi^2}
\left(\frac{e^{\gamma_E}}{8\pi}\right)^{-2\epsilon}
&&\left[(s^{-2\epsilon}+t^{-2\epsilon})\left(\frac{1}{4\epsilon^2}+\frac{a_1}{2\epsilon}\right)
+\frac{a_2}{\epsilon}\left(\frac{t^{-2\epsilon}s}{t}+\frac{s^{-2\epsilon}t}{s}\right)
\right. 
\\
&&\hspace{1mm}\left.
+a_3\left(\frac{s}{t}+\frac{t}{s}\right)-\frac{1}{2}\log^2\left(s/t\right)+a_4+\mathcal{O}(\epsilon)\right],
\label{DoubleBoxHard}
\eqae
where the coefficients are given by
\eqa
\nonumber a_1&=&1.528426,\\
\nonumber a_2&=&5/16,\\
\nonumber a_3&=&0.8224\pm 1.40\times 10^{-4}.\\
\nonumber a_4&=&-5.987\pm1.84\times 10^{-3},
\eqae

The remaining integrals are computed in appendix \ref{API0}. Adding everything together along with eq.(\ref{KiteResult}) and their $t$-channel counterparts, terms containing ratios $\frac{s}{t}$ and $\frac{t}{s}$ cancel, both in the divergent and the finite part, leaving behind a finite piece that is simply $\log^2\left(s/t\right)$ along with an additive constant. Note that the $\mathcal{O}(\epsilon^{-1})$ completely cancels as well. The final result is 
\eq
\framebox[14.2cm][c]{$\displaystyle\mathcal{A}_4^{2-loop}=\frac{-1}{16\pi^2}\left(\frac{N}{K}\right)^2\mathcal{A}_4^{tree}\left[\frac{(-\tilde \mu^2 s)^{-2\epsilon}}{(2\epsilon)^2}+\frac{(-\tilde \mu^2 t)^{-2\epsilon}}{(2\epsilon)^2}-\frac{1}{2}\log^2\left(\frac{-s}{-t}\right)+a+\mathcal{O}(\epsilon)\right]$}~,
\label{Final2}
\eqe
with $a=-8.021\pm 1.84\times 10^{-3}$ and $\tilde\mu\equiv \left(\frac{e^{\gamma_E}}{8\pi}\right)\mu$, where $\mu$ is the regularization constant.
\subsection{Evaluating $I_{0s}$ Using Five-Dimensional Formalism}
In the above derivation, $I_{0s}$ was computed in three-dimensional notations where dual conformal invariance of the integrand is not manifest. More precisely, the five-dimensional dual conformal tensor numerator in $I_{0s}$ breaks down to four conformally non-covariant numerators in three dimensions. One of the consequences is that we obtain a number of polynomials in $\frac{s}{t}$ and $\frac{t}{s}$ from some integrals which miraculously cancel at the very end.\footnote{Even if one chooses to use only scalar integrals via the identity in eq.(\ref{I0sIdentity}), one would again be left with four independent integrals.}  

From the one-loop computation in subsec.\ref{vanish1} one sees that it pays to maintain manifest dual conformal invariance in the intermediate steps. Indeed the one-loop tensor integral vanishes straight forwardly if one maintains the tensor numerator in five-dimensional notation, while in the three-dimensional notation the integral vanishes only upon the cancellation of two separate terms as shown in appendix \ref{vanish}. In the following we will redo the $I_{0s}$ integral and maintain five-dimensional notations in the intermediate steps to obtain the MB representation. Note that this approach is in spirit an extension of dimensional reduction, i.e. all tensor manipulations are done in five dimensions where dual conformal invariance is manifest and we only reduce to $D=3-2\e$ in the very end.

We again focus on the $l_1$ dependent part of the $I_{0s}$ integral, which in dual coordinates correspond to the $X_5$ dependent part. Instead of reducing to three-dimensional notations, we directly integrate away $X_5$ to obtain:
\beqa
\notag &&\int \mathcal{D}^3X_5 \frac{\epsilon(5,1,2,3,4)}{X_{51}^2X_{53}^2X_{54}^2X_{56}^2}\\
&&=\int^1_0 dF\frac{\Ga\left(-\frac{D}{2}+4\right)}{(4\pi)^{\frac{D}{2}}}\frac{\al_4\epsilon(6,1,2,3,4)}{ \left(\al_1\al_2 X_{13}^2+\al_1\al_4X_{16}^2+\al_2\al_4 X_{36}^2+\al_3\al_4 X_{46}^2\right)^{4-\frac{D}{2}}}.
\label{X5Part}
\eeqa
Note that although the $X_5$ integral is understood to be integrated on the projective light-cone in five-dimensions, we keep the dimension parameter $D$ in the result unfixed in anticipation of analytic continuation to $D=3-2\e$ at the very end. To rewrite the denominator as products of propagators, which will be useful for the $l_2$ integration, now $X_6$, we again convert the above integrand using the Mellin-Barnes representation. Integrating away the Feynman parameters, $I_{0s}$ becomes,
\beqa
I_{0s}=16t^{-1}Z_1\;A\;\int \mathcal{D}^3X_6 \left[\epsilon(6,1,2,3,4)\right]^2\left(X_{16}^2\right)^{z_1-1}\left(X_{36}^2\right)^{z_3-1}\left(X_{46}^2\right)^{z_2}\left(X_{26}^2\right)^{-1},
\eeqa
where $Z_1$ and $A$ are the same as in eqs.(\ref{Z1}) and (\ref{A}), respectively.
Once again the product of five-dimensional Levi-Civita tensors can be transformed into products of Kronecker deltas,
\beqa
16\epsilon(6,1,2,3,4)\epsilon(6,1,2,3,4)=st^2X_{61}^2X_{63}^2+s^2tX_{62}^2X_{64}^2.
\eeqa 
Through the above identity, $I_{0s}$ separates into two integrals, and the contribution of the second term from above identity gives us $I_{012}$  as in eq.(\ref{A12}) so that one arrives at the following MB representation:
\beqa
\notag I_{0s}&=&Z_1A\frac{st}{(X_{16}^2)^{-z_1}(X_{36}^2)^{-z_2}(X_{46}^2)^{-z_3}X_{26}^2}
+I_{012}\\
\notag &=&\int^{i\infty}_{-i\infty}\frac{dz_1dz_2dz_3dz_4}{(2\pi i)^4(4\pi)^{D}}
\frac{s^{-z_4+D-4}t^{z_4+1}\Ga(z_1+z_2+z_3-\frac{D}{2}+4)\Ga(z_1+z_2+z_3+2)}{\Ga(D-3)\Ga(z_1+z_2+z_3+D-1)}\\
\notag &&\times\textstyle \Ga(z_1+z_2-z_4+\frac{D}{2}-1)\Ga(z_2+z_3-z_4+\frac{D}{2}-1)\Ga(z_2+1)\Ga(z_4+1)\Ga(-z_2+z_4)\\
\notag &&\times \textstyle \Ga(-z_1-z_2-z_3+z_4-\frac{D}{2}+1)
\Ga(-z_1-z_2+\frac{D}{2}-3)\Ga(-z_2-z_3+\frac{D}{2}-3)\Ga(-z_4)\\
 &&+I_{012}~.
\eeqa

Explicit integration is again carried out using the $Mathemtica$ package MB.m~\cite{Czakon:2005rk}, the result for $I_{0s}$ plus the its $t$-channel counterpart is:
\beqa
\nonumber&& I_{0s}+(s\leftrightarrow t)=\frac{1}{16\pi^2}\left(\frac{e^{\gamma_E}}{8\pi}\right)^{-2\epsilon}\left[
(s^{-2\epsilon}+t^{-2\epsilon})\left(\frac{1}{4\epsilon^2}-\frac{1}{2\epsilon}\right)-\frac{1}{2}\log^2\left(s/t\right)+a'+\mathcal{O}(\epsilon)
\right],
\label{DoubleBoxHard2}
\eeqa
where $a'=-8.63480\pm 6.57\times 10^{-5}$. Combining eq.(\ref{DoubleBoxHard2}) with the result of the ``kite" integral along with its $t$-channel partner gives the same result as in eq.(\ref{Final2}).

\section{Discussions and Conclusions\label{guess}}
It is very interesting that the leading order quantum correction to the four-point amplitude of ABJM theory is the same as $\mathcal{N}=4$ super Yang-Mills. It is then natural to ask if the quantum corrections of the two theories can be mapped to each other beyond the present example, i.e. if the $2L$-loop correction of the four-point ABJM amplitude can be mapped to the $L$-loop correction of $\mathcal{N}=4$ super Yang-Mills amplitude.\footnote{The reason there is a factor of two difference can be understood via the transcendentality of loop amplitudes. } In this section we would like to present supporting arguments as to why this should be true at planar level. We also discuss in what form this duality between the two theories might take shape at higher points.

We note that there already exists evidence that the quantum corrections of the two theories are closely connected.\footnote{We would like to thank Radu Roiban and Gregory Korchemsky for detailed discussion of the cusp anomalous dimensions and infrared divergences of these two theories.} Indeed it was conjectured in~\cite{Gromov:2008qe}, and tested for in \cite{McLoughlin:2008ms,Alday:2008ut,Krishnan:2008zs,McLoughlin:2008he}, that the anomalous dimensions of twist operators of ABJM theory can be obtained from $\mathcal{N}=4$ super Yang-Mills, up to normalization factors which can be easily determined, simply by substituting, 
\eq
 \lambda_{\mathcal{N}=4} \rightarrow\left[4\pi h(\lambda_{ABJM})\right]^2,
 \label{replace}
 \eqe
where the $\lambda$s are the t'Hooft coupling of the two theories ($\lambda_{ABJM}=N/K$), and $h(\lambda)$ is a regularization scheme dependent function. At weak coupling $h(\lambda)$ has an expansion of the form~\cite{Gaiotto:2008cg}, 
\eq
h(\lambda)=\lambda(1+\sum_{i=1} c_i\lambda^{2i}+\cdot\cdot\cdot)\,. 
\eqe 
Since the anomalous dimension of the twist operators are related to the cusp anomalous dimension which partially controls the infrared divergence of the amplitude, combining this with the fact that the infrared singularity of the two-loop ABJM amplitude matches with the one-loop $\mathcal{N}=4$ super Yang-Mills result, and the property that infrared singularities exponentiate for the planar amplitudes~\cite{Sterman:2002qn,Gregory}, this suggests that the infrared singularity of the two theory, at four-point, share the same structure. However, the infrared singularity of $\mathcal{N}=4$ super Yang-Mills is controlled by both the cusp anomalous dimension and the collinear anomalous dimension. At one-loop the collinear anomalous dimension does not contribute in $\mathcal{N}=4$ super Yang-Mills, and hence whether the singularity structure of the two theories indeed matches depends on whether there is a similar correspondence between the collinear anomalous dimensions.

As mentioned in the introduction, for dual conformal invariant theories, the four-point amplitude can be uniquely determined by the anomalous dual conformal Ward-identiy. Our results indicate that the two-loop Ward-identiy of ABJM matches with the one-loop identity of $\mathcal{N}=4$ super Yang-Mills. Since the form of the anomalous Ward-identity is controlled by the structure of the infrared divergence, if the collinear anomalous dimension of the two theory matches, one concludes that the Ward-identity of the two theories continue to share the same structure at higher loops. This leads us to conjecture the following relationship between the four-point amplitudes of the two theories:
\eq
\left.\log\left( \mathcal{A}^{}_{4}/\mathcal{A}^{tree}_4\right)_{ABJM}=\quad\log\left( \mathcal{A}^{}_{4}/\mathcal{A}^{tree}_4\right)_{\mathcal{N}=4}\right|_{({\rm eq.}\;\ref{replace})}.
\eqe
Note that this would imply a BDS like ansatz for the four-point amplitude~\cite{Bern:2005iz}. This can be verified by an explicit four-loop computation.

There are obvious objections to Wilson loop/amplitude duality from strong coupling side. The trivial statement is that T-duality along three D2-brane world volume directions brings a IIA to a IIB theory. However, at weak coupling, the structure of the Yangian algebra~\cite{Bargheer:2010hn} and the coordinates of the dual space on which the dual superconformal generators act~\cite{Huang:2010qy} indicate that one should also T-dual three directions of the compact space. This would give even number of bosonic T-duality. Such a combination of fermionic T-duality was proposed for $AdS_5$/CFT$_4$ in~\cite{FermiT}, and its weak coupling representation was given in~\cite{Huang:2011um}. Therefore it would not be surprising that such 6$|$6 self-T-duality exists for $AdS_4$/CFT$_3$, although recent attempts at resolving this issue were complicated by the emergence of singularities~\cite{FermiT2}.

From the two-loop ABJM Wilson loop computation~\cite{Henn:2010ps}, one can easily see that the matter contribution will reproduce the same result as the one-loop $\mathcal{N}=4$ super Yang-Mills Wilson loop.\footnote{We would like to thank Konstantin Wiegandt for detailing the Wilson loop computation.} This is simply because the matter fields contribute in the form of an one-loop correction to the Chern-Simons gauge field propagator, resulting in the propagator taking the same form as the $\mathcal{N}=4$ super Yang-Mills gluon propagator. Thus the ABJM matter contribution will automatically reproduces the MHV $n$-point amplitude of $\mathcal{N}=4$ super Yang-Mills. Since the remaining pure Chern-Simons contribution can potentially provide additional dual conformal invariants, a natural guess at six-point is that the contribution is a sum of the additional dual-conformal invariants of the one-loop NMHV six-point amplitude, once the MHV piece is subtracted, i.e. the functions $V^{(i)}_{6}$ with $i=1,2,3$ defined in eq.(5.26) of~\cite{DualConformal0}.

On the amplitude side, based on the fermionic degree of the tree-level amplitudes of the two theories, one is tempted to conjecture that if one writes
\eqa
{\rm ABJM}:\quad\mathcal{A}^{2-loop}_{n}&=&\sum_{i}R_iM_i\\
{\rm \mathcal{N}=4 \;}:\quad\mathcal{A}^{1-loop}_{n}&=&\sum_{i}\tilde{R}_i\tilde{M}_i\qquad({\rm N}^{n/2-2}{\rm MHV}).
\eqae
where $R_i$ and $\tilde{R}_i$ are the dual conformal covariant spin structures appearing at tree level, and $M_i$ and $\tilde{M}_i$ are their quantum corrections respectively, then $M_i$ and $\tilde{M}_i$ are identical up to constants. As noted previously, the fact that the six-point tree level dual conformal covariant function is built out of the same fermionic functions could be viewed as a tacit support of this duality.   

One can understand the above conjecture as follows. If ABJM amplitudes were to somehow know about that of $\mathcal{N}=4$ sYM, then the fact that only even point amplitudes are non-vanishing should reflect some structure of $\mathcal{N}=4$ sYM. A natural suggestion is the fact that for $\mathcal{N}=4$ sYM, the lowest point for which ``honest"\footnote{Here by honest we mean that we only consider the lowest fermionic degree either in the chiral or anti-chiral representation. Hence a NMHV five-point amplitude will be categorized as an $\overline{\rm MHV}$ amplitude.} N$^K$MHV amplitudes begin to appear is when  $n=4+2K$. Furthermore, these $n$-point N$^{n/2-2}$MHV amplitudes are ``non-chiral", in the sense that they have the same fermionic degrees either in the chiral $SU(4)$ $\eta$ or anti-chiral $\bar{\eta}$ representation. This is parallel to the non-chiral property of ABJM amplitudes. A related fact is the property that the $n$-point ABJM tree-amplitude can be written in terms of ${n/2-2}$ products of dual conformal invariant fermionic functions, which is the same case for $n$-point N$^{n/2-2}$MHV $\mathcal{N}=4$ amplitudes. The dual conformal invariant fermionic functions take similar form as well, as demonstrated for the six-point amplitude in eq.(\ref{ABJMDCTree}). 

It may be curious why a Chern-Simons matter theory would know anything about four-dimensional helicity structures. This becomes plausible if one recalls that the four-dimensional helicity categorization as simply an expansion around the self-dual sector, whose S-matrix is trivial (at the quantum level for supersymmetric theories). There exists a trivial sector for Chern-Simons matter theory as well, i.e. the pure Chern-Simons part. Thus an $n$-point ABJM amplitude can be considered as  successive matter expansion around this trivial sector. In fact, for ABJM tree amplitude, using the first non-trivial four-point amplitude, one can construct an $n$-point amplitude by using recursion relations for $n/2-2$ times. This is the same structure as the CSW construction~\cite{Cachazo:2004kj} for N$^{n/2-2}$MHV amplitudes, where one starts with the first non-trivial amplitude, the MHV amplitude, and build the $n$-point amplitude via $n/2-2$ iterations. Alternatively, this can also be understood as the fermionic degrees in the dual conformal invariants.

The two-loop integrand is most conveniently expressed using tensor numerators are manifest dual conformal only in five dimensions. Indeed tensor integrals appears naturally from the view point of dual conformal symmetry and generally gives simpler integrand representation for amplitudes~\cite{Drummond:2010mb}. Here we demonstrate that maintaining manifest symmetry simplifies the calculation dramatically and obtains a final result where all the non-trivial cancellations seen in the conventional dimensional reduction approach becomes trivial.

Finally we like to comment that since the integrand were built from three-dimensional unitarity cuts, there are potential integrands that vanish in three dimensions and will not be detected by these cuts. One can obtain these terms by computing the unitarity cuts using the mass deformed tree amplitudes given in~\cite{Agarwal:2008pu}. However, as discussed in~\cite{Agarwal:2008pu}, the massive amplitudes written using massive three-dimensional spinor-helicity are essentially the same form as the massless case, one simply degenerates the complex spinors to real ones. Since our unitarity cut was checked analytically in spinor-helicity notations, the same integrand will agree with the massive unitarity cuts.

In summary, we have constructed the two-loop four-point integrand of ABJM theory utilizing the recently established dual conformal invariance and generalized unitarity methods. Explicit integration yields the same result as the four-cusp Wilson loop computation. This establishes the validity of the Wilson loop/amplitude duality for ABJM theory at two-loop four-point. Assuming that the duality holds beyond four-point implies a close relation between two-loop ABJM amplitudes and one-loop $\mathcal{N}=4$ super Yang-Mills amplitudes. 
\section{Acknowledgement}
The authors would especially like to thank Simon Caron-Hout for suggestions and verifications throughout this project. We would also like to thank Johannes Henn, Konstantin Wiegandt, Gregory Korchemsky, Radu Roiban, George Sterman and Zvi Bern for detailed discussion on various aspects of general loop amplitudes, as well as Marco S. Bianchi, Matias Leoni, Andrea Mauri, Silvia Penati and Alberto Santambrogio for private communications on the Feynman diagram calculation of the amplitude and pointing out a sign error in the early version of the draft. We would also like to thank Ilmo Sung, Tristan McLoughlin and Arthur Lipstein for discussion during the initial stage of this project. Part of this work was completed at the Kavli Institute for Theoretical Physics during the program ``The harmony of scattering amplitudes". WM thanks Professor Pei-Ming Ho and National Science Council of Taiwan to provide the exchange program to study in Caltech. The work of Y-t is supported by the US Department of Energy under contract DE-FG03-91ER40662 and in part by the National Science Foundation under Grant No. NSF PHY05-51164..
\appendix
\section{Conventions and Useful Formulas \label{AppA1}}
We follow the conventions used in ref.~\cite{Bargheer:2010hn}. The $SL(2,R)$ metric is
\eq
\epsilon_{\alpha\beta}=\left(\begin{array}{cc}~~0 & ~~1~ \\-1 & ~~0~\end{array}\right),\quad\epsilon^{\alpha\beta}=\left(\begin{array}{cc}~0 & -1~ \\~1 & ~~0~\end{array}\right).
\eqe
The spinor contraction is implemented as
\eq
\psi^\alpha \chi_\alpha =-\psi_\alpha \chi^\alpha, \quad \epsilon_{\beta\alpha}A^\alpha=A_\beta,\quad \epsilon^{\alpha\beta}A_{\beta}=A^{\alpha},\quad \epsilon^{\alpha\beta}\epsilon_{\beta\gamma}=\delta^\alpha_\gamma.
\eqe
The vector notation is translated to the bi-spinor notation
and vice versa through three-dimensional gamma matrices,
\eq
x^{\alpha\beta}=x^\mu (\sigma_\mu)^{\alpha\beta},\quad x^\mu=-\frac{1}{2}(\sigma^\mu)_{\alpha\beta}x^{\alpha\beta},
\eqe
with
\eq
\sigma^0=\left(\begin{array}{cc}-1 & ~~0~ \\~~0 & -1~\end{array}\right),\quad \sigma^1=\left(\begin{array}{cc}-1 & ~~0~ \\~~0 & ~~1~\end{array}\right),\quad \sigma^2=\left(\begin{array}{cc}~0 & ~~1~ \\~1 & ~~0~\end{array}\right),
\eqe
We list some useful identities
\begin{align}
&(\sigma^\mu)_{\alpha\beta}(\sigma^\nu)^{\alpha\beta}=-2\eta^{\mu\nu}\,,
\\
&(\sigma^\mu)_{\alpha\beta}(\sigma_\mu)_{\gamma\delta}=\epsilon_{\alpha\gamma}\epsilon_{\beta\delta}+\epsilon_{\beta\gamma}\epsilon_{\alpha\delta}\,,
\\
&\epsilon^{\mu\nu\rho}(\sigma_{\mu})^{ab}(\sigma_{\nu})^{cd}(\sigma_{\rho})^{ef}
=\frac{1}{2}(\epsilon^{ac}\epsilon^{be}\epsilon^{df}+\epsilon^{bc}\epsilon^{ae}\epsilon^{df}+\epsilon^{ad}\epsilon^{be}\epsilon^{cf}+\epsilon^{bd}\epsilon^{ae}\epsilon^{cf}
\nonumber \\
&\qquad\qquad\qquad\qquad\qquad\qquad
+\epsilon^{ac}\epsilon^{bf}\epsilon^{de}+\epsilon^{bc}\epsilon^{af}\epsilon^{de}+\epsilon^{ad}\epsilon^{bf}\epsilon^{ce}+\epsilon^{bd}\epsilon^{af}\epsilon^{ce}) \,,
\label{epsilon}
\\
& A^{[\alpha\beta]}=A^{\alpha\beta}-A^{\beta\alpha}=-\epsilon^{\alpha\beta}A^\gamma\,_\gamma,\\
& A_{[\alpha\beta]}=A_{\alpha\beta}-A_{\beta\alpha}=\epsilon_{\alpha\beta}A^\gamma\,_\gamma,\\
& x^{\alpha\beta}x_{\beta\gamma}=-x^2\delta^\alpha_\gamma,
\end{align}
where $x^2=x^\mu x_\mu$.\\
\par Here are some useful formulas for doing integration of $I_{0s}$ and $I_{1s}$:~\\
~\\
General Feynman parametrization,
\beqa
\label{FP}\left(\prod_{i=1}^{n} P_i^{\nu_i}\right)^{-1}=\Ga\left(\sum_{i=1}^n \nu_i\right)\int^1_0 \prod_{i=1}^n \frac{d\tau_i \tau_i^{\nu_i-1}}{\Ga(\nu_i)} \de\left(1-\sum_{i=1}^n \tau_i\right) 
\left(\prod_{i=1}^{n} \tau_i P_i^{\nu_i}\right)^{-\sum_{i=1}^n \nu_i}.
\eeqa
Integration of Feynman parameters,
\beqa
\label{IFP}\int^1_0 \prod_{i=1}^n d\tau_i\tau_i^{\si_i-1}  \de\left(1-\sum_{i=1}^n\tau_i\right)=\frac{\prod_{i=1}^n\Ga(\si_i)}{\Ga\left(\sum_{i=1}^n \si_i\right)}.
\eeqa

\section{Vanishing of $I_{4}^{1-loop}$\label{vanish}}
We we will show that the four-point one-loop integrand integrates to zero. We begin with the integrand, 
\eqa
\int \frac{d^D l_1}{(2\pi)^D}\frac{l_{1}^2\epsilon_{\mu\nu\rho}p_{1}^\mu p_2^\nu p_{4}^\rho+s\epsilon_{\mu\nu\rho}l_{1}^\mu p_{1}^\nu p_{4}^\rho}{l_1^2(l_1-p_1)^2(l_1-p_1-p_2)^2(l_1+p_4)^2}.
\eqae
Using Feynman parameters we arrive at, 
\eqa
\frac{1}{l_1^2(l_1-p_1)^2(l_1-p_1-p_2)^2(l_1+p_4)^2}
=\int_0^{1}dF \frac{\Ga(4) }{\left(l'^2_1+\alpha_2\alpha_4t+\alpha_3\alpha_1s\right)^4},
\eqae
where, 
\beqa
l_1'&=&l_1-\alpha_2p_1+\alpha_3p_3+(1-\alpha_1-\alpha_2)p_4.
\label{shift2}
\eeqa
Now since the denominator contains only $l'^2_1$, terms in the numerator with only linear to $l'^\mu$ will integrate to zero, and the above integral can be further simplified to, 
\eqa
\nonumber&&\Ga(4)\int_0^{1}dF\int \frac{d^Dl'_1}{(2\pi)^D}  \frac{(l_1'^2-\alpha_2\alpha_4t-\alpha_3\alpha_1s)\epsilon_{\mu\nu\rho}p_{1}^\mu p_2^\nu p_{4}^\rho}{\left(l'^2_1+\alpha_2\alpha_4t+\alpha_3\alpha_1s\right)^4}\\
&&=\Ga(4)\int_0^{1}dF\int \frac{d^Dl'_1}{(2\pi)^D} \left[-\frac{\epsilon_{\mu\nu\rho}p_{1}^\mu p_2^\nu p_{4}^\rho}{\left(l'^2_1+\alpha_2\alpha_4t+\alpha_3\alpha_1s\right)^3} +2\frac{l_1'^2\epsilon_{\mu\nu\rho}p_{1}^\mu p_2^\nu p_{4}^\rho}{\left(l'^2_1+\alpha_2\alpha_4t+\alpha_3\alpha_1s\right)^4}\right].
\eqae
Performing the $l'$ integration gives:
\eqa
\epsilon_{\mu\nu\rho}p_{1}^\mu p_2^\nu p_{4}^\rho\int_0^{1}dF \frac{\Gamma(-\frac{D}{2}+3)}{(4\pi)^{D/2}\Gamma(3)}\left(-3+D\right)(\alpha_2\alpha_4t+\alpha_3\alpha_1s)^{3-\frac{D}{2}}.
\eqae
Taking $D=3-2\epsilon$, the above vanishes. 

\section{Integrals of $I_{0s}$}\label{API0}
Here we complete the integrals which have not been shown in section \ref{I0s}. 
One can follow the same steps with Feynman parameterization and MB representation as in section \ref{I0s} to show
\beqa
\notag &&\int \frac{d^D l_2}{(2\pi)^D}\frac{1}{[(l_2)^2]^{\si_1}[(l_2+p_4)^2]^{\si_2}[(l_2-p_1-p_2)^2]^{\si_3}}\\
\label{IA11}&=&\frac{\Ga\left(\mbox{$\si_1+\si_2+\si_3-\frac{D}{2}$}\right)\Ga\left(\mbox{$-\si_2-\si_3+\frac{D}{2}$}\right)\Ga\left(\mbox{$-\si_1-\si_2+\frac{D}{2}$}\right)s^{-\si_1-\si_2-\si_3+\frac{D}{2}}}{(4\pi)^{\frac{D}{2}}\Ga(\si_1)\Ga(\si_3)\Ga\left(\mbox{$-\si_1-\si_2-\si_3+D$}\right)},\\
\notag &&\int \frac{d^D l_2}{(2\pi)^D}\frac{1}{\left[(l_2)^2\right]^{\si_1}\left[(l_2+p_4)^2\right]^{\si_2}\left[(l_2-p_1-p_2)^2\right]^{\si_3}\left[(l_2-p_1)^2\right]^{\si_4}}\\
\notag &=&\int_{-i\infty}^{+i\infty}\frac{dz_4}{2\pi i}\frac{\Ga(-z_4)
\Ga\left(\mbox{$-\si_2-\si_3-\si_4-z_4+\frac{D}{2}$}\right)\Ga(\si_2+z_4)t^{z_4}s^{-\si_1-\si_2-\si_3-\si_4-z_4+\frac{D}{2}}}{(4\pi)^{\frac{D}{2}}\Ga(\si_1)\Ga(\si_2)\Ga(\si_3)\Ga(\si_4)\Ga\left(\mbox{$-\si_1-\si_2-\si_3-\si_4+D$}\right)}\\
\label{IA12}&&\times\Ga\left(\mbox{$\si_1+\si_2+\si_3+\si_4+z_4-\frac{D}{2}$}\right)\Ga\left(\mbox{$-\si_1-\si_2-\si_4-z_4+\frac{D}{2}$}\right)\Ga(\si_4+z_4).
\eeqa
One can substitute some specific $\si_i$s into eqs.(\ref{IA11}) and (\ref{IA12}) which represent the correspondent integrals in eqs.(\ref{A12})-(\ref{A15}).

The remaining term of eq.(\ref{I01}), $I_{01B}$, can be easily obtained along the the same line as we go through above.
Let us now consider $I_{02}$ and $I_{03}$. In fact, $I_{02}=I_{03}$ because if we interchange $k_1\leftrightarrow k_4$, $k_2\leftrightarrow k_3$ and $l_1\leftrightarrow -l_2$ which will interchange expressions of $I_{02}$ and $I_{03}$ without changing the Mandelstam variables. Moreover, the integration of $l_1$ in $I_{01}$, $I_{02}$ and $I_{03}$ are the same. We find,

\beqa
\notag I_{012}&=&\int^{+i\infty}_{-i\infty} \frac{dz_1dz_2dz_3}{(4\pi)^D(2\pi i)^3}
\frac{s^{D-3}\Ga(z_1+z_2+z_3-\frac{D}{2}+4)\Ga(z_1+z_2+z_3+2)}{\Ga(D-3)\Ga(z_1+z_2+z_3+D-1)\Ga(-z_1+1)\Ga(-z_3+1)}\\
\notag &&\times\textstyle  \Ga(z_1+z_2+\frac{D}{2})\Ga(z_2+z_3+\frac{D}{2})\Ga(z_2+1)
 \Ga(-z_1-z_2-z_3-\frac{D}{2}+1)\\
&&\times\textstyle\Ga(-z_1-z_2+\frac{D}{2}-3)\Ga(-z_2-z_3+\frac{D}{2}-3)\Ga(-z_1)\Ga(-z_2)\Ga(-z_3),\\
 \notag I_{013}&=& -\int^{+i\infty}_{-i\infty} \frac{dz_1dz_2dz_3}{(4\pi)^D(2\pi i)^3}
 \frac{s^{D-3}\Ga(z_1+z_2+z_3-\frac{D}{2}+4) \Ga(z_1+z_2+z_3+2)}{\Ga(D-3)\Ga(z_1+z_2+z_3+D-1)\Ga(-z_3+1)}\\
\notag&&\textstyle\times\Ga(z_1+z_2+\frac{D}{2})\Ga(z_2+z_3+\frac{D}{2}-1)\Ga(z_2+1)\Ga(-z_1-z_2-z_3-\frac{D}{2}+1) \\
&&\textstyle\times\Ga(-z_1-z_2+\frac{D}{2}-3)\Ga(-z_2-z_3+\frac{D}{2}-3)\Ga(-z_2)\Ga(-z_3),\\
\notag  I_{014}&=&-\int^{+i\infty}_{-i\infty} 
\frac{dz_1dz_2dz_3dz_4}{(4\pi)^D(2\pi i)^4}\frac{s^{-z_4+D-3}t^{z_4}\Ga(z_1+z_2+z_3-\frac{D}{2}+4)\Ga(z_1+z_2+z_3+2)}{\Ga(D-3)\Ga(z_1+z_2+z_3+D-1)\Ga(-z_2-1)\Ga(-z_3+1)}\\
\notag&&\textstyle \times \Ga(z_1+z_2-z_4+\frac{D}{2})\Ga(z_2+z_3-z_4+\frac{D}{2}-1)\Ga(z_2+1)\Ga(z_4+1)\\
\notag&&\times \textstyle \Ga(-z_1-z_2-z_3+z_4-\frac{D}{2}+1)\Ga(-z_1-z_2+\frac{D}{2}-3)\Ga(-z_2-z_3+\frac{D}{2}-3) \\
&&\times\Ga(-z_2+z_4-1)\Ga(-z_2)\Ga(-z_3)\Ga(-z_4),\\
\notag  I_{015}&=&\int^{+i\infty}_{-i\infty} 
\frac{dz_1dz_2dz_3dz_4}{(4\pi)^D(2\pi i)^4}\frac{s^{-z_4+D-3}t^{z_4}\Ga(z_1+z_2+z_3-\frac{D}{2}+4)\Ga(z_1+z_2+z_3+2)}{\Ga(D-3)\Ga(z_1+z_2+z_3+D-1) \Ga(-z_1-1)\Ga(-z_3+1)}\\
\notag&&\textstyle\times\Ga(z_1+z_2-z_4+\frac{D}{2})\Ga(z_2+z_3-z_4+\frac{D}{2}-2)\Ga(z_2+1)\Ga(z_4+1)\\
\notag&&\textstyle \times \Ga(-z_1-z_2-z_3+z_4-\frac{D}{2}+1)\Ga(-z_1-z_2+\frac{D}{2}-3)\Ga(-z_2-z_3+\frac{D}{2}-3)
\\
&&\textstyle \times \Ga(-z_2+z_4)\Ga(-z_1)\Ga(-z_3)\Ga(-z_4),\\
\notag  I_{01B}&=&\left(1+\frac{t}{s}\right)\int^{+i\infty}_{-i\infty} 
\frac{dz_1dz_2dz_3dz_4}{(4\pi)^D (2\pi i)^4}
\frac{s^{-z_4+D-3}t^{z_4}\Ga(z_1+z_2+z_3-\frac{D}{2}+4)\Ga(z_1+z_2+z_3+1)
}{\Ga(D-3)\Ga(z_1+z_2+z_3+D-2)\Ga(-z_1+1)\Ga(-z_3+1)}\\
\notag&&\textstyle\times \Ga(z_1+z_2-z_4+\frac{D}{2}-1)\Ga(z_2+z_3-z_4+\frac{D}{2}-2)\Ga(z_2+1)\Ga(z_4+1)\\
\notag&&\textstyle\times \Ga(-z_1-z_2-z_3+z_4-\frac{D}{2}+3)\Ga(-z_1-z_2+\frac{D}{2}-2)\Ga(-z_2-z_3+\frac{D}{2}-3)\\
&&\textstyle\times\Ga(-z_2+z_4)\Ga(-z_1)\Ga(-z_3)\Ga(-z_4),
\eeqa 
\beqa
\notag I_{02}&=&
-\left(1+\frac{t}{s}\right)\int^{+i\infty}_{-i\infty} 
\frac{dz_1dz_2dz_3dz_4}{(4\pi)^D(2\pi i)^4}
\frac{ s^{-z_4+D-3}t^{z_4}\Ga(z_1+z_2+z_3-\frac{D}{2}+4)
}{ \Ga(D-3)\Ga(1-z_3)}\\
\notag&&\times\textstyle\Ga(z_2+z_3-z_4+\frac{D}{2}-2)\Ga(z_2+1)\Ga(z_4+1)\Ga(-z_1-z_2-z_3+z_4-\frac{D}{2}+2)\\
\notag &&\times\textstyle\Ga(-z_2-z_3+\frac{D}{2}-3\Big)\Ga(-z_2+z_4)\Ga(-z_3)\Ga(-z_4)\\
&&\times\left[\frac{ \Ga(z_1+z_2+z_3+2)\Ga(-z_1-z_2+\frac{D}{2}-3)\Ga(z_1+z_2-z_4+\frac{D}{2})}{\Ga(z_1+z_2+z_3+D-1)}\right.\\
\notag &&\hspace{2mm}~~+
\left.\frac{ \Ga(z_1+z_2+z_3+1)\Ga(-z_1-z_2+\frac{D}{2}-2)\Ga(z_1+z_2-z_4+\frac{D}{2}-1)}{\Ga(z_1+z_2+z_3+D-2)}\right],\\
\notag I_{03}&=&I_{02},
\eeqa
\beqa
\notag I_{04}&=&
\int^{+i\infty}_{-i\infty}\frac{dz_1}{(4\pi)^{D}(2\pi i)}
 \frac{(s^2t^{D-5}+st^{D-4})\Ga(-D+5)\Ga(z_1-\frac{D}{2}+3)}{\Ga(D-3)\Ga(\frac{3D}{2}-5)\Ga(z_1-\frac{D}{2}+3)}\\
  &&\times \textstyle\Ga(z_1+\frac{D}{2}-2)\Ga(z_1+1)\Ga(-z_1+\frac{D}{2}-2)\Ga(\frac{D}{2}-2)\Ga(-z_1)
    \Ga(D-4)\,.
\eeqa   
Explicit integration gives:
\beqa
 I_{012}&=&\frac{1}{32\pi^2}\left(\frac{e^{\gamma_E}s}{8\pi}\right)^{-2\epsilon}\left[-\frac{b_1}{2\e}+b_2+\mathcal{O}(\epsilon)\right],\\
\notag&&(b_1=1.306853,~b_2=-1.98584\pm 9.29\times 10^{-5}) \\
 I_{013}&=&\frac{1}{32\pi^2}\left(\frac{e^{\gamma_E}s}{8\pi}\right)^{-2\epsilon}\left[-\frac{1}{2\e}+c_1+\mathcal{O}(\epsilon)\right],\\
\notag&&(c_1=0.19315\pm 8.37\times 10^{-5})\\
 I_{014}&=&\frac{1}{32\pi^2}\left[-\frac{d_1}{2\epsilon}\left(\frac{e^{\gamma_E}s}{8\pi}\right)^{-2\epsilon}+\frac{d_2}{2\epsilon}\frac{s}{t}\left(\frac{e^{\gamma_E}t}{8\pi}\right)^{-2\epsilon}+d_3\frac{s}{t}+d_4+\mathcal{O}(\epsilon)\right],\\
\notag&&(d_1=1/2,~d_2=1/2,~d_3=-1.84657\pm 2.70\times 10^{-5}, d_4=-0.460279)\\
 I_{015}&=&\frac{1}{32\pi^2}\left(\frac{e^{\gamma_E}t}{8\pi}\right)^{-2\epsilon}\left[-\frac{1}{2\e}\left(e_1\frac{s^2}{t^2}+e_2\frac{s}{t}\right)+e_3\frac{s^2}{t^2}+e_4\frac{s}{t}+\mathcal{O}(\epsilon)\right],\\
\notag&&(e_1=5/4,~ e_2=3/4,~ e_3=0.105140\pm 2.62\times 10^{-6},~e_4=0.70171\pm 1.14\times10^{-5})\\
  I_{01B}&=&\frac{1}{32\pi^2}\left(1+\frac{t}{s}\right)\left(\frac{e^{\gamma_E}t}{8\pi}\right)^{-2\epsilon}\left[-\frac{1}{2\e}\left(f_1\frac{s^2}{t^2}+f_2\frac{s}{t}\right)+f_3\frac{s^2}{t^2}+f_4\frac{s}{t}+\mathcal{O}(\epsilon)\right],\\
\notag&&(f_1=-5/4, ~f_2=1/4,~ f_3=-1.105140\pm 2.63\times10^{-6},~f_4=2.99143\pm2.25\times10^{-5})\\
  I_{02}&=&\frac{1}{32\pi^2}\left(1+\frac{t}{s}\right)\left(\frac{e^{\gamma_E}t}{8\pi}\right)^{-2\epsilon}\left[-\frac{1}{2\e}\frac{s}{t}+g_1\frac{s^2}{t^2}+g_2\frac{s}{t}+\mathcal{O}(\epsilon)\right],\\
\notag&&(g_1=1\pm 5.30\times10^{-5}, ~g_2=-1.69314\pm 6.95\times10^{-5})\\
\notag  I_{03}&=& I_{02},\\
  I_{04}&=&-\frac{1}{32\pi^2}\left(\frac{s^2}{t^2}+\frac{s}{t}\right).
\eeqa

The final answer consists of summing the above result along with eq.(\ref{KiteResult}), combining this with their $t$-channel counter parts (by exchanging $s\leftrightarrow t$), and adding this to eq.(\ref{DoubleBoxHard}). One then arrives at eq.(\ref{Final2}).


\end{document}